\documentclass[5p]{elsarticle}

\usepackage{url}
\usepackage{color}
\usepackage{booktabs}
\usepackage{amsfonts}      
\usepackage{rotating}

\journal{Reliability Engineering \& Safety Systems}

\newif\ifcomments\commentstrue

\ifcomments
\newcommand{\authornt}[3]{\textcolor{#1}{[#3 ---#2]}}
\newcommand{\todo}[1]{\textcolor{red}{[TODO: #1]}}
\else
\newcommand{\authornt}[3]{}
\newcommand{\todo}[1]{}
\fi

\newcommand{\colAwidth}{0.2\columnwidth}
\newcommand{\colBwidth}{0.755\columnwidth}

\begin{document}

\begin{frontmatter}

\title{Building Confidence in Scientific Computing Software Via Assurance Cases}

\author{Spencer~Smith}
\ead{smiths@mcmaster.ca}
\author{Mojdeh Sayari Nejad}
\author{Alan Wassyng}
\ead{wassyng@mcmaster.ca}
\address{Computing and Software Department, McMaster University. 1280 Main Street West,
  Hamilton, Ontario, Canada, L8S 4K1}

\begin{abstract}
  Assurance cases provide an organized and explicit argument for
  correctness. They can dramatically improve the certification of Scientific
  Computing Software (SCS). Assurance cases have already been effectively used
  for safety cases for real time systems. Their advantages for SCS include
  engaging domain experts, producing only necessary documentation, and providing
  evidence that can be verified/replicated. This paper illustrates assurance
  cases for SCS through the correctness case for 3dfim+, an existing Medical
  Imaging Application (MIA) for analyzing activity in the brain. This example
  was partly chosen because of recent concerns about the validity of fMRI
  (Functional Magnetic Resonance Imaging) studies. The example justifies the
  value of assurance cases for SCS, since the existing documentation is shown to
  have ambiguities and omissions, such as an incompletely defined ranking
  function and missing details on the coordinate system. A serious concern for
  3dfim+ is identified: running the software does not produce any warning about
  the necessity of using data that matches the parametric statistical model
  employed for the correlation calculations. Raising the bar for SCS in general,
  and MIA in particular, is both feasible and necessary – when software impacts
  safety, an assurance case methodology (or an equivalently rigorous confidence
  building methodology) should be employed.
\end{abstract}

\begin{keyword}
assurance cases \sep software quality \sep software requirements
  specification \sep scientific computing \sep medical imaging software
\end{keyword}

\end{frontmatter}

\section{Introduction} \label{SecIntroduction}

Are we currently putting too much trust in the quality of Scientific Computing
Software (SCS)?  For instance, for Functional Magnetic Resonance Imaging (fMRI),
concerns exist with respect to the statistical analysis models commonly
employed~\cite{Guardian2016, PNAS2016}.  Since medical professionals use the
output of fMRI, and other Medical Imaging Applications (MIA), for diagnosis and
treatment planning, we need to have confidence in the software.  Often the
developers of SCS, such as MIA, are medical physicists, scientists and
engineers, not software engineers.  Although SCS developers do excellent work,
are there currently enough checks and balances, from a software development
perspective, for confidence in correctness?  The usual approach employed when
correctness is critical is to impose requirements for official software
certification, where the goal for certification is to: ``...systematically
determine, based on the principles of science, engineering and measurement
theory, whether a software product satisfies accepted, well-defined and
measurable criteria'' \cite[p.~12]{HatcliffEtAl2009}.  Unfortunately, five
significant problems exist for SCS in general, and MIA in particular, in
completing a conventional certification exercise through an external body, like
the Food and Drug Administration (FDA):

\begin{enumerate}
\item The external body cannot rely on all their staff having deep expertise in
  the physical problem the software simulates or analyses, or in the numerical
  techniques employed.  This tends to lead the external body to request a large
  quantity of documentation, as shown in standards for medical
  software~\cite{CDRH2002}.
\item SCS developers tend to dislike documentation.  Scientists do not view
  rigid, process-heavy approaches, favourably~\cite{CarverEtAl2007}.  Moreover,
  they often consider reports for each stage of software development as
  counterproductive~\cite[p.~373]{Roache1998}.  Although documentation is
  generated when required, the normal work flow has the documentation as a
  ``necessary evil'' at the end of the process.
\item Historically when software engineers work with scientists there are
  challenges for communication and collaboration~\cite{Kelly2007, Segal2005}.
\item Conventional documentation (requirements, design, etc.) relies on an
  implicit argument for correctness; a tacit assumption is made that completing
  the documentation will improve quality, but no explicit argument is given to
  show that correctness will be a consequence of the documentation.
  Perhaps more important, there is no explicit recognition that adequate and effective
  documentation is necessary for correctness of complex software applications.
\item Verification of SCS is challenging because of the oracle
  problem~\cite{KellyEtAl2011} -- testing is difficult because we do not always
  know the expected correct output for a given set of inputs.
\end{enumerate}

A potential solution to these problems is to have the SCS developers create, or
partially create, an assurance case before, or while they develop their
software.  Assurance case techniques have been developed and successfully
applied for real time safety critical systems~\cite{RinehartEtAl2015,
  Rushby2015, Wassyng2015}.  An assurance case presents an organized and
explicit argument for correctness (or whatever other software quality is deemed
important) through a series of sub-arguments and evidence.  Putting the argument
in the hands of the experts means that they will work to convince themselves,
along with the regulators.  They will use the expertise that the regulators may
not have; they will be engaged.  This engagement will hopefully help bridge the
current chasm between software engineering and scientific
computing~\cite{Kelly2007, Storer2017}, by motivating scientists toward
documentation and correcting the problem of software engineers failing to meet
scientists' expectations~\cite{Segal2008}.  Significant documentation will still
likely be necessary, but through assurance cases the developers now decide the
content of the documentation.  What is created will be relevant and necessary.
More details on the current literature on assurance cases is given in
Section~\ref{SecAssuranceCases}.

Arguing in favour of assurance cases for SCS does not imply that SCS developers
have not, or do not currently treat correctness seriously.  They have developed
many successful theories, techniques, testing procedures and review processes.
In fact, an assurance case will likely use much of the same evidence that SCS
developers currently use to convince themselves of the correctness of their
software.  The difference is that the argument will no longer be ad hoc, or
incompletely documented.  The argument will now be explicitly presented for
review by third parties; we will no longer be implicitly asked to trust the
developer.  The act of creating the assurance case may also lead the developer
to discover subtle edge cases, which would not have been noticed with a less
rigorous and systematic approach.  This is particularly true when testing is
complicated by the lack of a test oracle.  The developer needs to overcome this
challenge and their solution to the problem should be open to external scrutiny.

While the eventual goal is to develop a template for assurance cases for any
SCS, our initial approach is to learn by first building an assurance case for
one particular example.  We ask ourselves what an assurance case would look like
for an MIA example, and then assess the potential value of this assurance case.
Our case study focuses on 3dfim+, an MIA software package that supports
Functional Magnetic Resonance Imaging (fMRI).  3dfim+ was selected because it is
a reasonably small (approximately 1700 lines of code) and easy to understand
example of medical image analysis software.  3dfim+ also has the advantage that
testing is straightforward, because independent implementations of the
calculations exist that provide a pseudo-oracle.  We targeted medical image
analysis software, since some of the common fMRI statistical analyses data have
not yet been validated~\cite{Guardian2016} and because a recent
study~\cite{PNAS2016} has shown a potentially serious flaw in software commonly
used to analyze fMRI data.  More detail on 3dfim+ can be found in
Section~\ref{Sec3dfim+}.

The scope of our work does not include redeveloping or reimplementing
3dfim+. Our goal is to build an assurance case for the existing software by
treating it as black box.  We consider only the executable for 3dfim+ and the
existing documentation. We produce new documentation and testing results, but
not new code.  Excluding the code makes the case study more realistic, since, if
an assurance case exercise were to be conducted in industry, there would be
little appetite for reimplementation.  Considerable effort has already gone into
writing medical image analysis (and other scientific software); it is not
feasible for the community to start over.

To argue for the correctness of 3dfim+, we developed an assurance case with the
top claim of `` Program 3dfim+ delivers correct outputs when used for its
intended purpose in its intended environment, and within its assumed operating
assumptions.''  Part of the explicit argument involved developing a Software
Requirements Specification (SRS) document that contains all the necessary
information and mathematical background needed to understand 3dfim+.  This
document can be used for validation and verification activities; sections of it
appear many times as evidence in our assurance case. The SRS was reviewed by a
domain expert as part of the assurance case evidence.  We also developed a test
case to illustrate how the results from 3dfim+ can be checked to provide
additional evidence of correctness.  An early version of the full assurance case
for 3dfim+ can be found in Nejad 2017~\cite[Appendix B]{Nejad2017}.  Excerpts
from the full case are given in Section~\ref{SecCaseStudy}.  A brief overview of
this work is provided in Smith et al 2018~\cite{SmithEtAl2018_ICSEPoster}.

Besides providing a means to illustrate assurance cases for MIA, the 3dfim+
example provides an opportunity to justify the value of assurance cases for
certification.  Although no errors were found in the output of the existing
software, the rigour of the proposed approach did lead to discovering
ambiguities and omissions in the existing documentation.  Moreover, the
assurance case highlighted a potential safety concern when running the software
itself.  Most importantly, the explicit arguments and artifacts included in the
assurance case provide evidence that can be independently judged for
sufficiency.  The specific evidence for the validation of assurance cases for
MIA is given in Section~\ref{SecValidation}, while the generalization of the
approach for other SCS applications is presented in
Section~\ref{SecGeneralization}.

\section{Overview of Assurance Cases} \label{SecAssuranceCases}

An assurance case is ``[a] documented body of evidence that provides a convincing
and valid argument that a specified set of critical claims about a system's
properties are adequately justified for a given application in a given
environment''~\cite[p.\@ 5]{RinehartEtAl2015}.

The idea of assurance cases (or safety cases) began after a number of serious
accidents, starting with the Windscale Nuclear Accident in the late 1950s.  This
incident was the United Kingdom's most serious nuclear power accident
\cite{ONR2016} and was instrumental in the government setting up new safety
regulations incorporating assurance cases.  Although there had not previously
been an ignorance of safety concerns, and safety standards and regulatory
approaches had been applied as the norm, the previous approaches proved to be
insufficient.  They lacked interaction between regulators and developers,
especially since the developers were often more knowledgeable than the
regulators about the safety of their products.  Assurance cases (safety cases)
do not just focus on verifying and validating the parts, but also on the
interaction between the parts that may cause something unexpected to emerge.

Assurance cases have been widely used in the European safety community for over
20 years to ensure system safety~\cite{Kelly99}. They have been applied in
industries such as aerospace, transportation, nuclear power, and
defence~\cite{Bishop98}.  Other examples include the energy sector, aviation
infrastructure, aerospace vehicles, railways, automobiles, and medical devices,
such as pacemakers, and infusion pumps~\cite{RinehartEtAl2015}. Attempts have
also been made to develop assurance cases in the security
sectors~\cite{Charles2007}.

In North America, the medical domain is showing an increased interest in
assurance cases.  Safety cases are considered to have ``the potential to support
healthcare organizations in the implementation of structured and transparent
systems for patient safety management''~\cite{HealthFoundation2012}.  This
potential is reflected in the Food and Drug Administration's (FDA's) strong
recommendation that manufacturers submit a safety assurance case for any new
infusion pumps~\cite{FDA2014}.

Safety cases, and in general assurance cases, require a clearly articulated
argument, supported by evidence.  An assurance case consists of a claim that we
make about the properties of a product, that is then supported by sub-claims
that are eventually grounded in evidence derived from the product itself and its
development.

\begin{figure}[!h]
\centering
\includegraphics[width=\columnwidth]{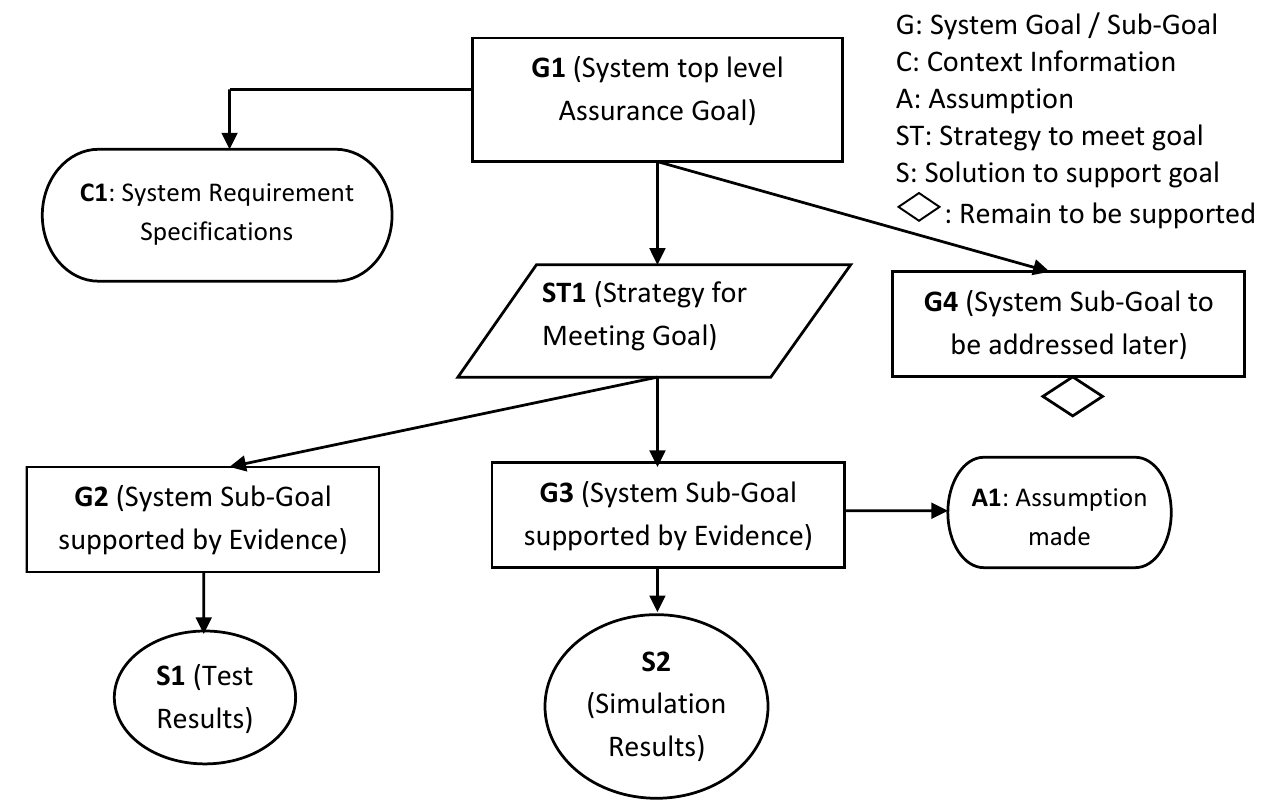}
\caption{A basic GSN structure \cite{GadeAndDeshpande2015} }
\label{element}
\end{figure}

For our work we have chosen the popular Goal Structuring Notation (GSN),
developed by Kelly~\cite{Kelly2003} to make our arguments clear, easy to read
and, hence, easy to challenge. To develop the assurance case, we used Astah
(\url{http://astah.net/}) to create and edit our GSN assurance cases.  GSN
starts with a Top Goal (Claim) that is then decomposed into Sub-Goals, and
terminal Sub-Goals are supported by Solutions (Evidence). Strategies describe
the rationale for decomposing a Goal or Sub-Goal into more detailed
Sub-Goals. There are other constructs in GSN; a full overview, with examples,
can be found in Sprigg's book~\cite{Spriggs2012}.  Figure~\ref{element} shows
what an assurance case might look like, using Goal Structuring Notation (GSN)
for goals, context and assumptions.

Focusing on the assurance case from the start of a project can improve the
efficiency of the development process. SCS, such as MIA, is often subject to
standardization and regulatory approval. While applying such approvals and
standards has had a beneficial effect on system quality, it does not provide
good tracking of the development stages, as the compliance with the standards
are mostly checked after the system development. Once a system is implemented,
its documentations must be approved by the regulators. This process is lengthy
and expensive. In contrast, assurance case development should progress at least
in parallel with the system construction (as recommended by the
FDA~\cite{FDA2014}), resulting in a traceable, detailed argument for the desired
property (or properties). Moreover, assurance cases take a more direct, flexible
and explicit approach. They are flexible enough to incorporate all existing
assurance activities and artifacts in any step of the procedure.  With the aid
of a template and with experience, we believe that developing an assurance case
does not necessarily require as much additional effort as people fear, and it
potentially reduces costs, saves time and gives greater freedom in accommodating
different standards.

\section{Overview of 3dfim+} \label{Sec3dfim+}

\begin{figure}[!h]
\centering
\includegraphics[width=1.0\columnwidth]{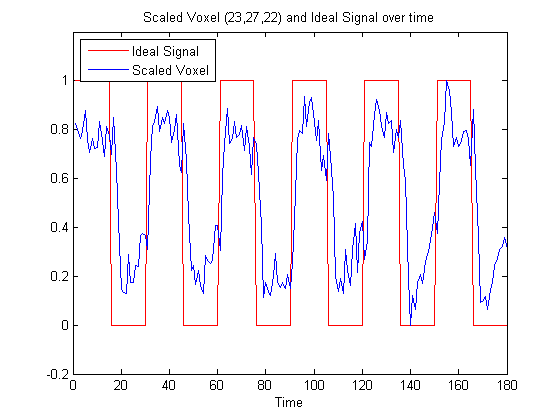}
\caption{Ideal signal versus activity of the voxel at position (23,27,22) over time}
\label{idealvsbrain}
\end{figure}

3dfim+~\cite{Ward2000} is a tool in the Analysis of Functional NeuroImages
(AFNI) package (\url{https://afni.nimh.nih.gov/}).
3dfim+ analyzes the activity of the brain by computing the correlation between
an ideal signal and the measured brain signal.  The ideal signal is defined by
the user.  For instance, the ideal signal could be a square wave, as shown in
Figure~\ref{idealvsbrain}.  For ease of comparison, the corresponding value of
the measured activity in Figure~\ref{idealvsbrain} is scaled between 0 and 1.
This figures shows high correlation between ideal and measured signals.

\begin{figure}[!h]
\centering
\includegraphics[width=0.8\linewidth]{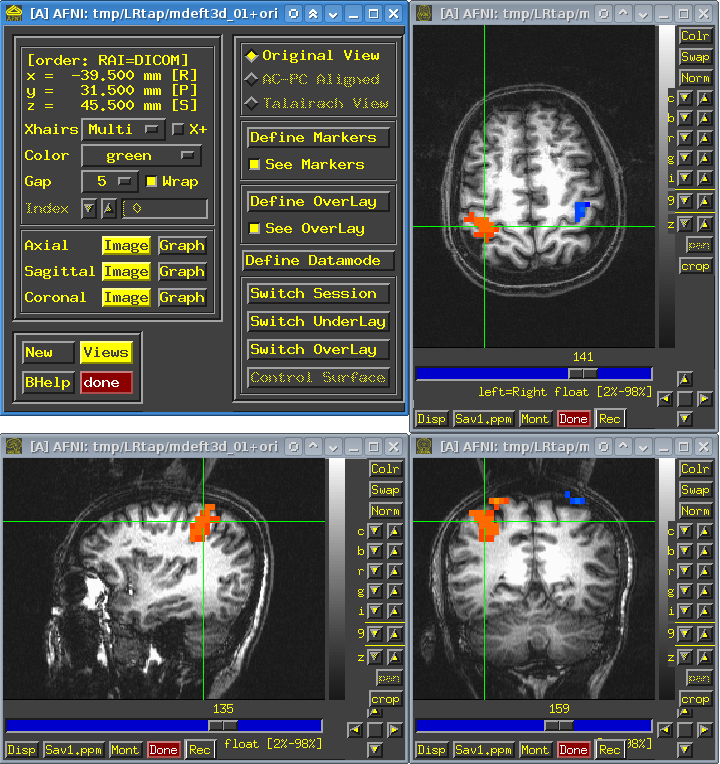}
\caption{AFNI environment and visualizing the active parts of the brain (from
  \texttt{https:// commons.wikimedia.org/ wiki/ File: AFNI\_screenshot.png})}
\label{AFNI}
\end{figure}

Figure~\ref{idealvsbrain} shows the ideal signal versus the brain activity for
one voxel in the full 3D image of the brain.  This analysis is completed for
every voxel.  The results can be visualized using the tools in the AFNI.
Figure~\ref{AFNI} shows the AFNI environment, in which we can see the brain from
different perspectives with areas of high (negative and positive) correlation
highlighted.

As mentioned in Section~\ref{SecAssuranceCases}, assurance cases are usually
developed in parallel with the system construction.  Given that 3dfim+ already
exists, this was not an option for our current case study.  This means that we
had to be particularly vigilant to avoid problems with confirmation bias.  We
did not want to prove correctness of 3dfim+ with a flawed argument.  Since we do
not have a vested interest in the correctness of 3dfim+, this is less likely to
be a problem than it might generally be.  Our initial ignorance of the domain
area also helps, since we acquired the domain knowledge in parallel with
constructing the assurance case.

\section{Assurance Case for 3dfim+} \label{SecCaseStudy}

We have used the guidance provided in ``General Principles of Software
Validation; Final Guidance for Industry and FDA Staff''~\cite{GPFDA} to develop
our assurance case. This guide outlines generally recognized validation
principles that are FDA acceptable for the medical software validation. It was
prepared by the International Medical Device Regulators Forum (IMDRF) to provide
globally harmonized principles concerning medical device software. The general
principles document includes software, like 3dfim+, that is itself considered a
medical device.

The presentation of the assurance case starts with an overview of assurance
arguments using GSN.  This is followed by summaries and excerpts from the
evidence used to support the argument.  The evidence includes the Software
Requirements Specification (SRS), Test Cases, and Domain Expert Review.

\subsection{Assurance Case}

Our assurance case consists of many sub-claims, which means it cannot be
represented legibly on a single page.  Therefore, we will only include a
representative subset of the argument, which has been split to separately show
the sub-structures.

We have to label all parts of the assurance case structure, i.e. all goals,
evidence, and contexts, so that our arguments can be discussed and reviewed
unambiguously. A number of strategies exist to do this~\cite[p.\
32--33]{Spriggs2012}. For the ease of navigation, we prefer a hierarchical
scheme; top goals in each sub-structure are labeled with a word or a letter but
without a number (for example G) and then their sub-goals are labeled as G.1,
G.2, ... and the subgoals of G.1 and G.2 are labeled, respectively, as G.1.1,
G.1.2, ... and G.2.1 and G.2.2, ... and so on. The evidence is labeled in a
similar way. Contexts, strategies, evidence and justifications are labeled
alphabetically if more than one context, strategy or justification is used for
an argument; for example, C\_Ga, C\_Gb, C\_Gc, ... for contexts and S\_Ga,
S\_Gb, S\_Gc for strategies of the Goal G and so on.

When splitting a goal into its sub-goals, the rationale behind the choice of
sub-goals is explained using strategies.  In cases where the rationale is
straightforward, it is excluded for space consideration.  We have confined
ourselves to traditional GSN, although we believe that GSN should be augmented
to include \emph{reasoning}, not just a \emph{strategy} for decomposition. The
original intent for assurance cases was to make the argument explicit. Without
the reasoning that demonstrates that sub-goals act as premises for their parent
goal, the argument remains implicit.

\begin{figure*}[!h]
\centering
\includegraphics[width=0.8\linewidth]{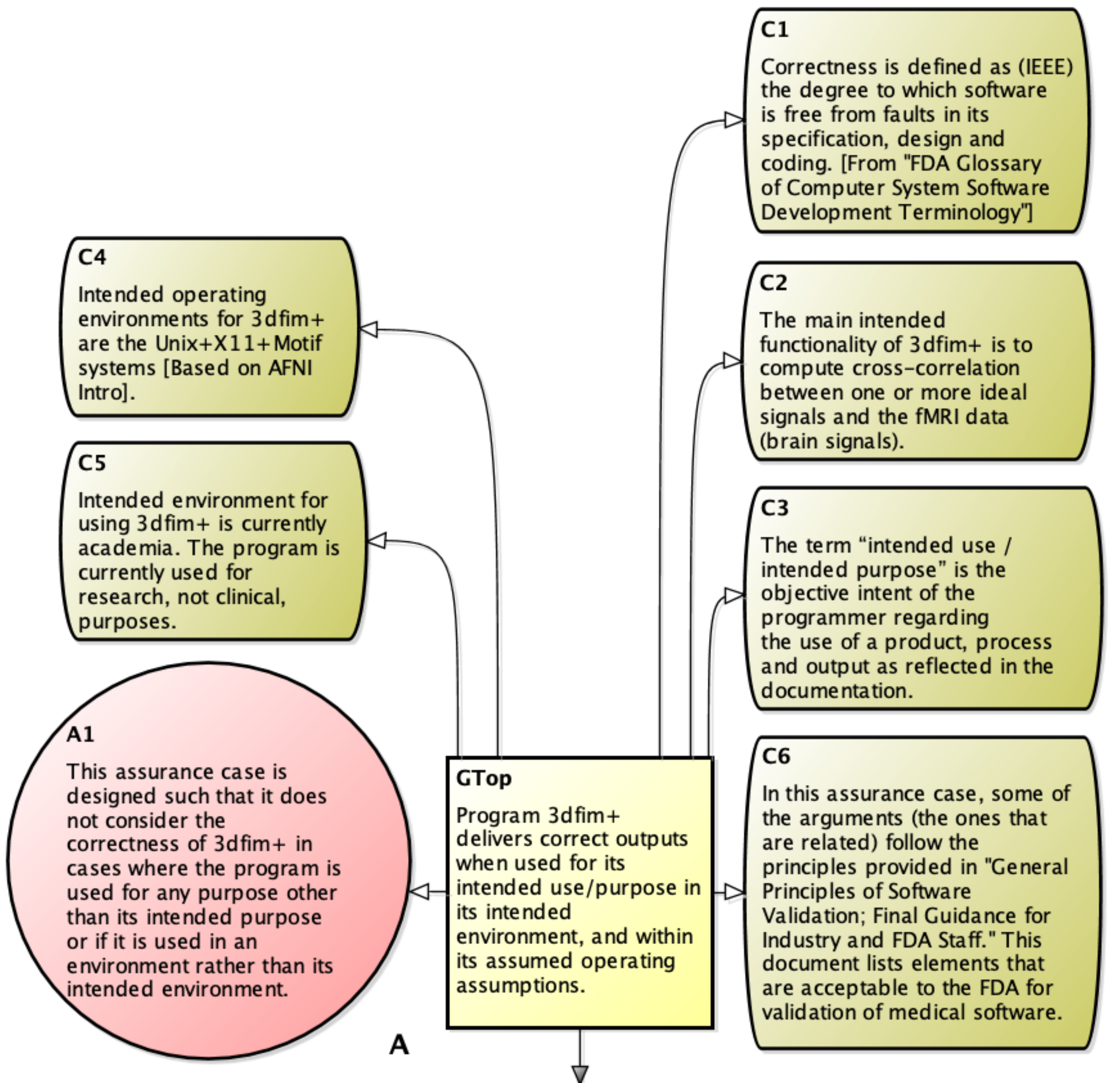}
\caption{Contexts and Assumption in Top Goal}
\label{TopAC}
\end{figure*}

We have defined our top goal as ``Program 3dfim+ delivers correct outputs when
used for its intended use/purpose in its intended environment, and within its
assumed operating assumptions.''  The truth of a claim depends on its context;
therefore, we must be explicit about what we mean by each term in our goal
statement. We could include the details with the goal statement itself, but then
it would be too long and would lose its focus.  The solution is to declare the
context separately. We have defined each term in the top goal in several
contexts. We have also made an assumption that the 3dfim+ will only be used for
its intended purpose in its intended environment. The assumption and contexts
are shown in Figure~\ref{TopAC}.

\begin{figure*}[!h]
\centering
\includegraphics[width=0.8\linewidth]{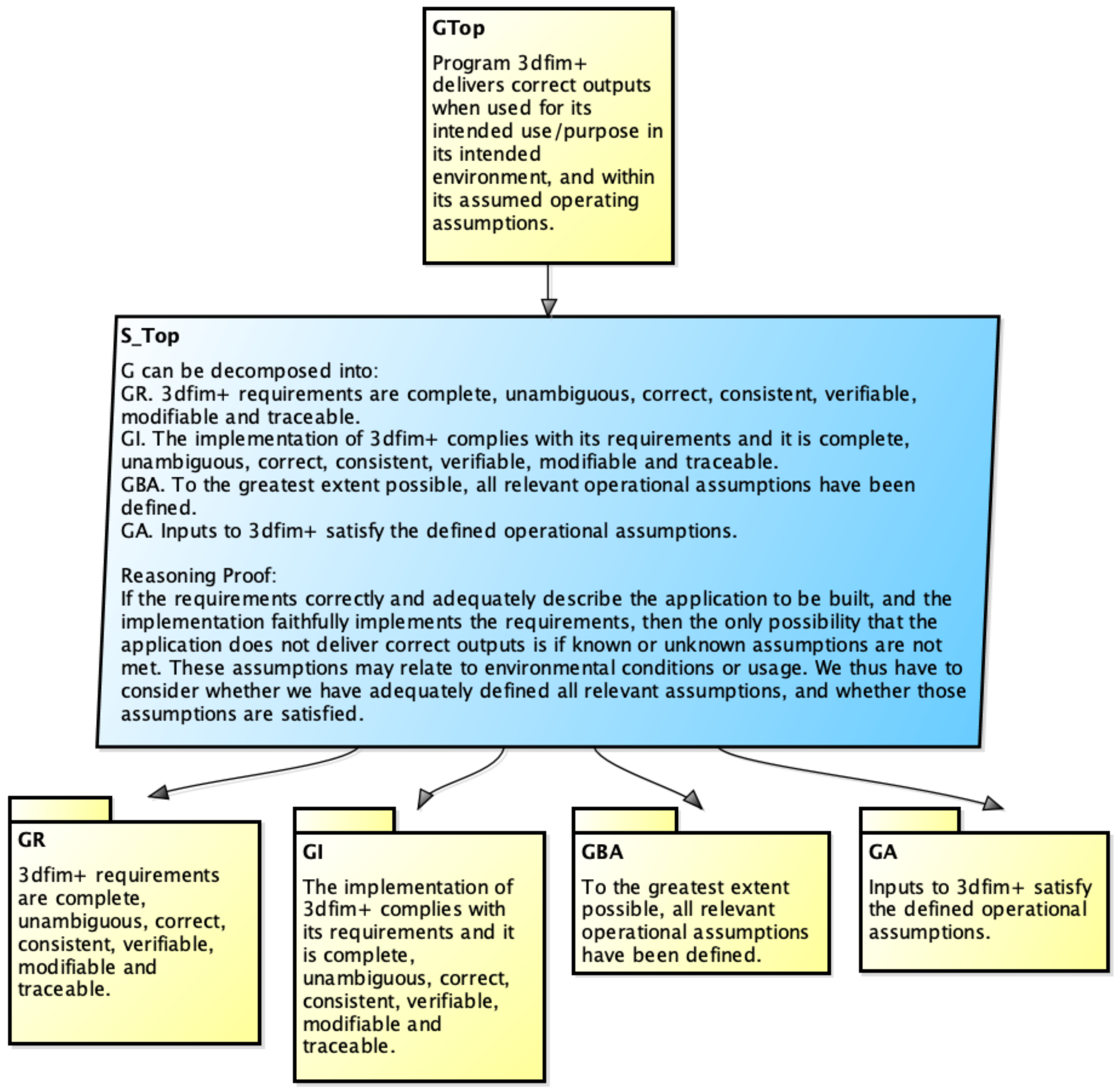}
\caption{Top Goal of the assurance case and its sub-goals}
\label{TopGoal}
\end{figure*}

As previously done for medical device assurance cases~\cite{Wassyng2015}, we
have divided the top goal into four sub-goals, as shown in Figure~\ref{TopGoal}.
The first sub-goal (GR) argues for the quality of the documentation of the
requirements.  To make an overall argument for correctness, we need a
specification to judge correctness against.  The second sub-goal (GI) says that
the implementation complies with the requirements and the third sub-goal (GBA)
states that, to the extent possible, the relevant operational assumptions have
been identified.  The fourth sub-goal GA also relates to assumptions, claiming
that the inputs to 3dfim+ will satisfy the operational assumptions; we
need valid input to make an argument for the correctness of the output.  The
strategy section of the GSN diagram presents the reasoning for decomposing the
top-goal in this way:  ``If the requirements correctly and adequately describe the
application to be built, and the implementation faithfully implements the
requirements, then the only possibility that the application does not deliver
correct outputs is if known or unknown assumptions are not met. These
assumptions may relate to environmental conditions or usage. We thus have to
consider whether we have adequately defined all relevant assumptions, and
whether those assumptions are satisfied.''

The top level of the assurance case in Figure~\ref{TopGoal} does not imply that
up-front requirements are needed.  This is fortunate, since scientists have the
view that requirements are impossible to determine up-front, since they believe
that details can only emerge as the work progresses~\cite{CarverEtAl2007,
  SegalAndMorris2008}.  The assurance case needs requirements, but they can come
out of the development process in any way appropriate for the developers.  That
is, the documentation can be ``faked'' like it is part of a rational design
process~\cite{ParnasAndClements1986}.

\begin{figure*}[!h]
\centering
\includegraphics[width=0.7\linewidth]{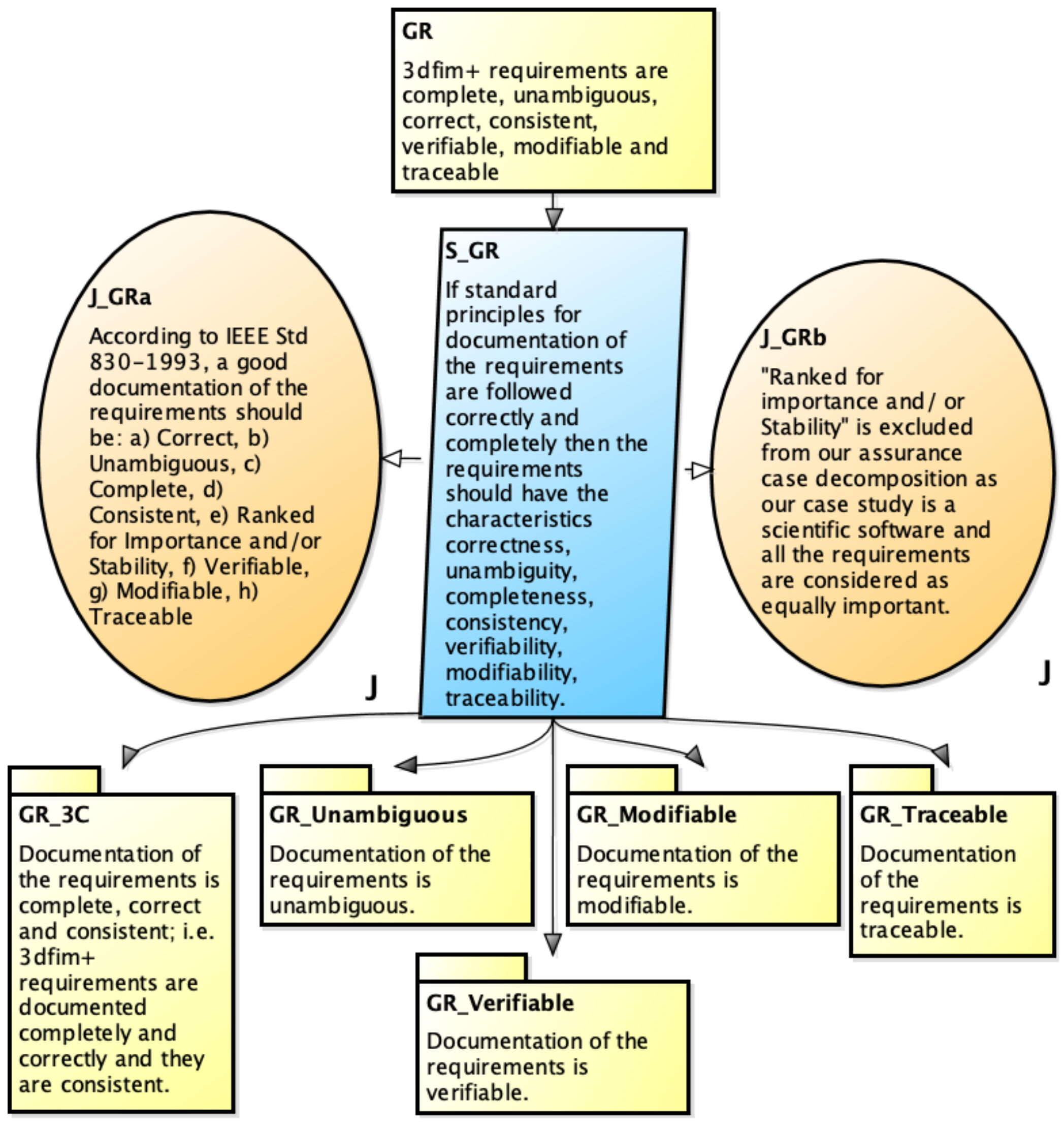}
\caption{GR decomposition}
\label{GRTop}
\end{figure*}

The main focus in our assurance case is arguing for GR (quality requirements).
The decomposition of GR into its sub-goals is shown in Figure~\ref{GRTop}. This
decomposition is based on the IEEE standard 830-1993~\cite{IEEESRS93}. This
standard states that good documentation of requirements should be correct,
unambiguous, complete, consistent, ranked for importance and/or stability,
verifiable, modifiable and traceable (J\_GRa).  Using the IEEE resource
increases confidence in the argument and makes it more compelling.  Our
sub-goals address correctness, unambiguity, completeness, consistency,
verifiability, modifiability and traceability of the requirements documentation.
``Ranked for importance and/or stability'' is excluded from the sub-goals in
Figure~\ref{GRTop} (as shown in J\_GRb) because our domain is MIA, where for the
software to function properly, all requirements are considered of equal
importance.  This is shown as justification J\_GRb in Figure~\ref{GRTop}.

The arguments for consistency, completeness, and correctness were combined
together in goal G\_3C. These qualities were grouped because, according to some
publications, such as `` The Three Cs of Requirements: consistency,
completeness, and correctness''~\cite{ThreeCs}, there is an important
relationship between completeness, consistency and correctness for software
requirements. Improving one of these three qualities may diminish the
others. From another perspective, correctness is a combination of consistency
and completeness. So it is important to consider these 3 qualities together.
The argument for completeness is partially based on the argument for the
readiness of a business plan from Spriggs~\cite[p.\ 30]{Spriggs2012}.  Due to
space limitations, the full details of this argument are not included here.
They can be found in~\cite{Nejad2017}.

\begin{sidewaysfigure*}[htpb]
\centering
\includegraphics[width=0.85\textwidth]{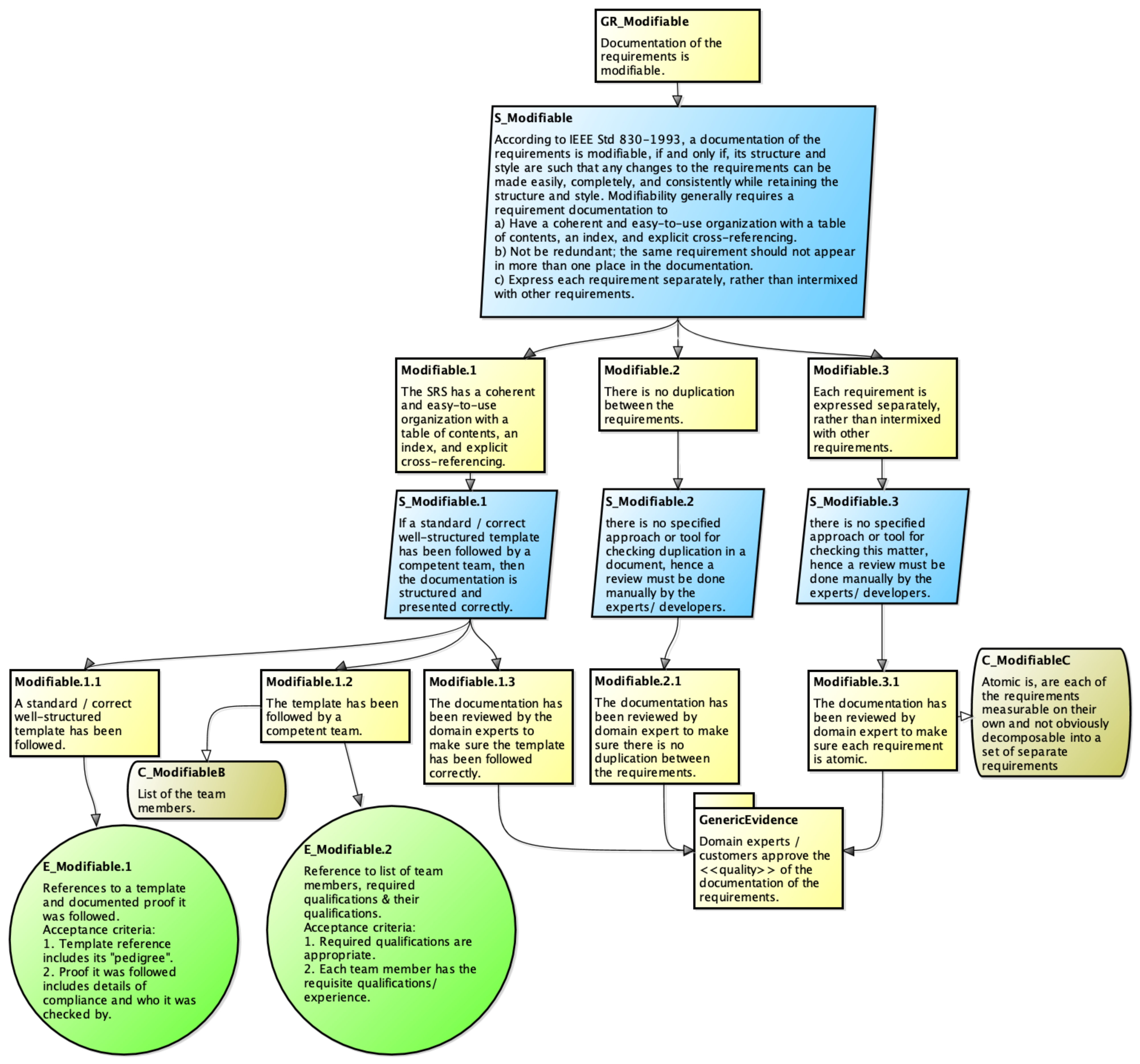}
\caption{Argument for modifiability of documentation of requirements}
\label{Modifiable}
\end{sidewaysfigure*}

\begin{figure*}[htpb]
\centering
\includegraphics[width=1.05\textwidth]{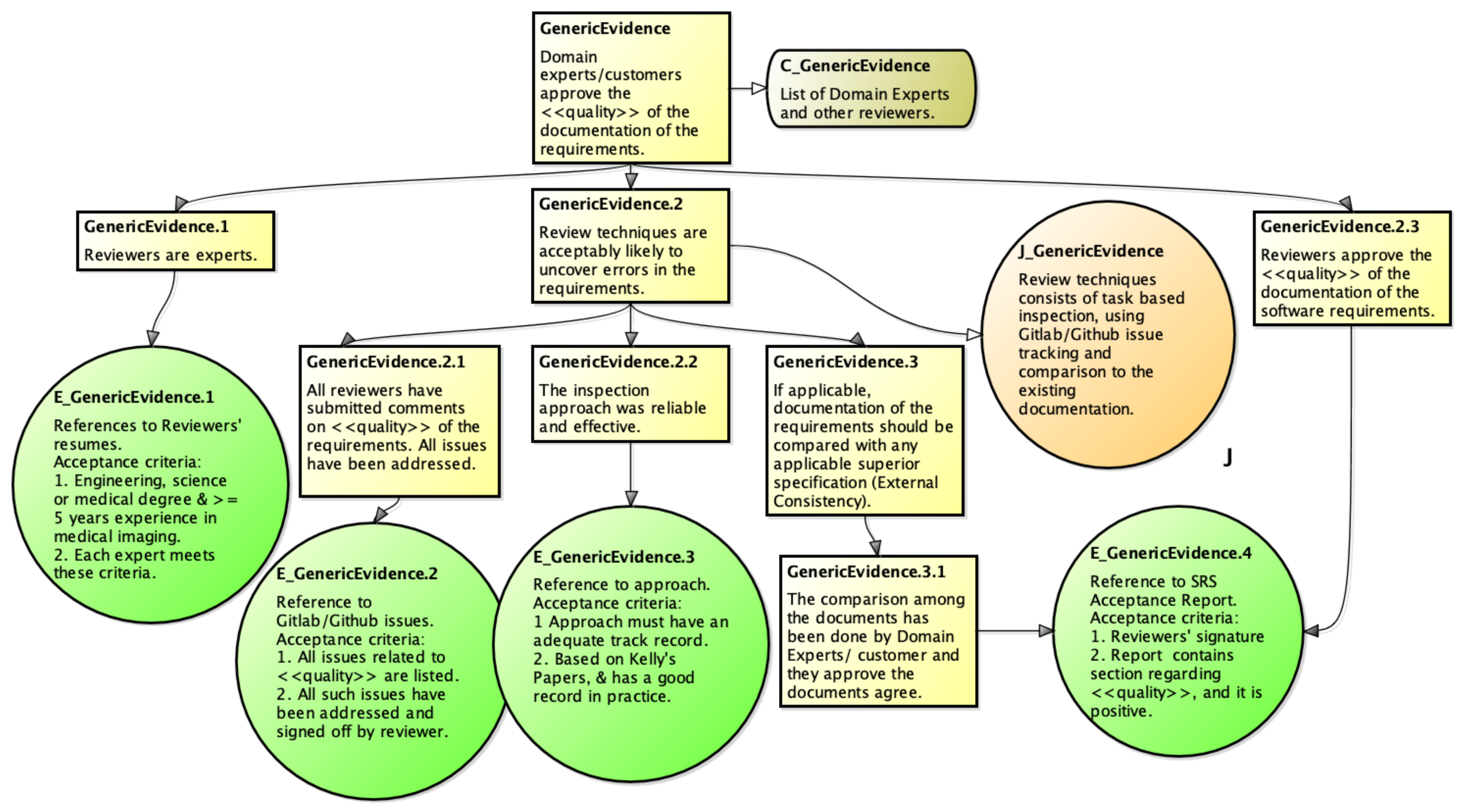}
\caption{Generic evidence module used as a pattern in our assurance case}
\label{GenericEvidence}
\end{figure*}

A sample expansion for the sub-goal of modifiability from GR
(Figure~\ref{GRTop}) is shown in Figure~\ref{Modifiable}.  Modifiability is a
quality attribute of the software architecture that relates to ``the cost of
change and refers to the ease with which a software system can accommodate
changes''~\cite{Northrop2004}.  Modifiability generally requires the
requirements documentation to have a coherent and easy-to-use organization with
a table of contents, an index, and explicit cross-referencing. Moreover,
requirements should not be redundant and they must be expressed separately.  As
for the other qualities, the argument for modifiability makes use of the generic
evidence template (Figure~\ref{GenericEvidence}) discussed below.

The content of the documentation of the requirements must be reviewed and
verified by domain experts. This is particularly important in SCS because of the
special role of assumptions.  We cannot hope to develop models that include all
of the complexities of the real world, so the adopted simplifying assumptions
need to be judged for appropriateness.  Spriggs~\cite[p.\ 37]{Spriggs2012} gives
a decomposition for arguments that end with domain expert review.  We have
developed a similar decomposition in our template modules, called
GenericEvidence, as shown Figure~\ref{GenericEvidence}.  GenericEvidence is a
generic argument. The generic argument is often called a ``pattern''. ``A
pattern in this context is an argument that applies to a class of things, which
you can use as the basis of an argument for a specific instance''~\cite[p.\
103]{Spriggs2012}.  We have developed this module to re-use it for several
arguments in our assurance case. We have an argument that a particular quality
of the requirements documentation has been met; the main evidence items are the
acceptance report and the addressed comments submitted by the reviewers. If we
want to ensure that another quality has been met, we do not want to start our
argument again from scratch. We prefer to use the same module (sub-structure),
but bring in a new evaluation, comments and sections in the report as
evidence. In that case, we can have the name of the quality in the module, but
publish the argument stating exactly which quality is reviewed. For instance,
for the sake of completeness, we verified that all statements made in the
original documentation are reflected in the new documentation. This comparison
is mentioned as GenericEvidence.3 in Figure~\ref{GenericEvidence}.

E\_GenericEvidence.1 in Figure~\ref{GenericEvidence} mentions the acceptance
criteria for reviewers' resumes.  This information is included in the assurance
case to mitigate against the bias problem mentioned in
Section~\ref{SecAssuranceCases}.  Reviewers need to be qualified, and we should
say what qualified means before we start looking for a reviewer.  When we
document the assurance case, we verify that the experts satisfy these criteria.
If not, we have to make an argument why they should still be considered experts.
This draws attention to the fact that there is something ``unusual'' here that
may typically be overlooked.

\begin{figure*}[!h]
\centering
\includegraphics[width=0.7\linewidth]{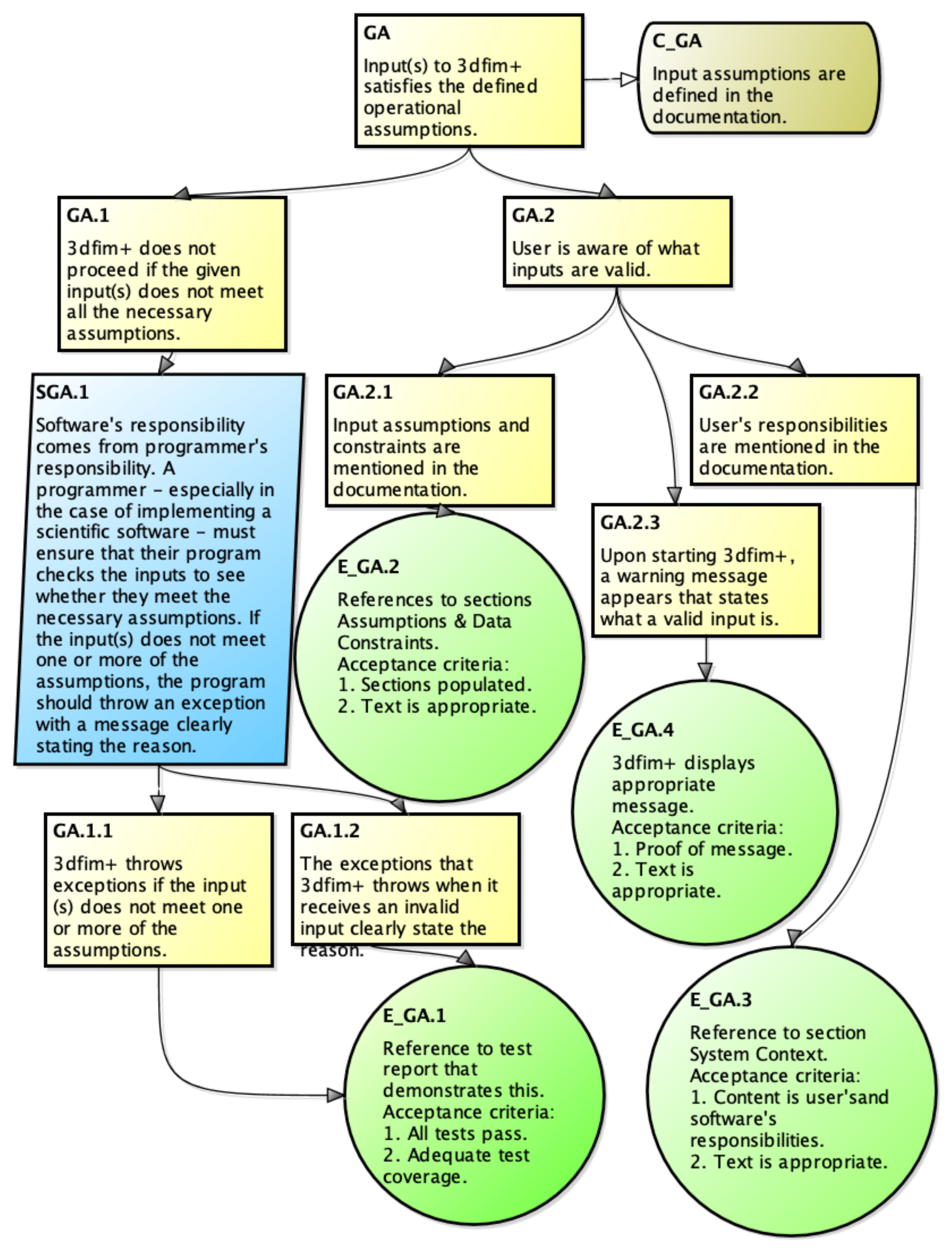}
\caption{Argument for inputs satisfying the defined operational assumptions}
\label{AssumptionGoal}
\end{figure*}

In Figure~\ref{TopGoal} we defined GA as ``Inputs to 3dfim+ satisfies the
defined operational assumptions.''  Achieving this goal partially relies on the
software to check if the input is valid, but not all inputs can be validated by
the software.  The user of 3dfim+ also has responsibility, in the same sense
that an automobile driver has responsibility to operate their vehicle safely.
This argument for GA is shown in Figure~\ref{AssumptionGoal}.  This argument
explicitly states that the user has responsibility for validating the input.
The software can do automated checks, like verify that the measured activities
are positive, but the software can never tell if 3dfim+ is the right tool for
the job.  For instance, the statistical model for 3dfim+ is parametric, if a
non-parametric model would be more appropriate, the user will have to select
another tool.  Although not currently part of 3dfim+, we added a warning
message, as part of the assurance case, that users be explicitly reminded of
their responsibilities while running the software, similar to how movies or
video games with flashing lights warn of the possibility of triggering seizures.
Argument GA demonstrates the value of assurance cases requiring a complete
argument.  The responsibility of the user can easily be forgotten without an
explicit coverage requirement to check that no cases are missing.

\subsection{Software Requirements Specification}

Having a Software Requirements Specification (SRS) is critical for software
validation~\cite{GPFDA}.  The requirements mentioned in the assurance case for
goal GR (Figure~\ref{GRTop}) are documented in the SRS.  This document is
necessary to verify software correctness, since it provides a specification
against which correctness can be judged.  As a consequence, the SRS is mentioned
in the sub-goals and evidence for several goals.  For instance, in
Figure~\ref{Modifiable}, for modifiability, the SRS is mentioned in goal
Modifiable.1.  Figure~\ref{GenericEvidence} for generic evidence references the
SRS in E\_Generic Evidence.4, by calling for a requirements acceptance report.
Goal GA (Figure~\ref{AssumptionGoal}) for the operational assumptions imposes
several requirements on the SRS, such as in E\_GA.2, which mentions SRS content
related to assumptions and data constraints.

Due to space limitations, only some representative excerpts from the SRS for
3dfim+ can be reproduced here.  The excerpts selected are intended to give an
overall feel for the document and to highlight some areas where the rigour of
the assurance case provides benefits for the documentation and software
quality.  The full SRS is available in~\cite[Appendix A]{Nejad2017}.  

\subsubsection{SRS Template} \label{SecTemplate}

Writing an SRS generally starts with a template, which provides guidelines and
rules for documenting the requirements.  The assurance case supports the need
for a template through the modifiability goal (Figure~\ref{Modifiable})
Modifiable.1.1: ``A standard/correct well-structured template has been
followed.''  Several existing templates contain suggestions on how to avoid
complications and how to achieve qualities such as verifiability,
maintainability and reusability~\cite{ESA1991, IEEE1998, NASA1989}.  However, no
template is universally accepted.  For the MIA example, the choice was a
template specifically designed for scientific software~\cite{SmithAndLai2005,
  SmithEtAl2007}, as illustrated via the table of contents shown below.  The
recommended template is suitable for science, because of its hierarchical
structure, which decomposes abstract goals to concrete instance models, through
the support of data definitions, assumptions and terminology.  The document's
structure facilitates its maintenance and reuse~\cite{SmithAndLai2005}, by using
separation of concerns, abstraction and traceability.

\begin{enumerate}
\item Reference Material
\begin{enumerate}
\item Table of Units
\item Table of Notations
\item Table of Symbols
\item Abbreviations and Acronyms
\end{enumerate}
\item Introduction
\begin{enumerate}
\item Purpose of Document
\item Scope of Requirements
\item Organization of Document
\end{enumerate}
\item General System Description
\begin{enumerate}
\item System Context
\item User Characteristics
\item System Constraints
\end{enumerate}
\item Specific System Description
\begin{enumerate}
\item Problem Description
\begin{enumerate}
\item Background
\item Terminology Definition
\item Coordinate Systems
\item Physical System Description
\item Goal Statements
\end{enumerate}
\item Solution Characteristics Specification
\begin{enumerate}
\item Assumptions
\item Theoretical Models
\item Data Definitions
\item Instance Models
\item Data Constraints
\item Properties of a Correct Solution
\end{enumerate}
\end{enumerate}
\item Requirements
\begin{enumerate}
\item Functional Requirements
\item Non-functional Requirements
\end{enumerate}
\item Other System Issues
\item Traceability Matrix
\item Likely Changes
\end{enumerate}

\subsubsection{Goals}

The high level objectives of the software are documented in the goals (Section
4.a.v of the SRS template shown in Section~\ref{SecTemplate}).  A sample goal
for 3dfim+ is:

\begin{itemize}

\item[\textbf{G1:}] Estimate the Pearson correlation coefficients between the
  (best) ideal time series and the fMRI time series at each voxel over time.
 
\end {itemize}

\subsubsection{Assumptions}

Assumptions (Section 4.b.i of the SRS template) highlight a simplification made
for the purpose of the mathematical modelling.  A significant responsibility of
the SRS is to document the assumptions.  As mentioned above, making
the assumptions explicit facilitates expert review.  Sample assumptions for
3dfim+ include:

\begin{itemize}

\item[\textbf{A1:}] The variables should be either of type interval or ratio.

\item[\textbf{A2:}] There is a linear relationship between the two variables.
 
\item[\textbf{A3:}] The variables are bivariately normally distributed.

\end {itemize}

\subsubsection{Theoretical Models} \label{SecTheoryModel}

The theoretical models are sets of governing equations or axioms that are used
to model the problem described in the problem definition section (SRS Section
4.b.ii).  Traceability exists between the theoretical model and the other components of
the documentation.  For instance, the description for the T1 (Pearson
Correlation Coefficient), which is given below, references the definition for mean (DD1) and several
assumptions, including the three listed above.

~\newline
 \noindent
\begin{minipage}{\columnwidth}
\begin{tabular}{@{} p{\colAwidth}  p{\colBwidth}@{}}
\toprule
{Number} & \textbf{T1} \\
\midrule
{Label} &\bf Calculating Pearson Correlation Coefficient\\
\midrule
Equation&  $\rho(A,B) = \frac{ \sum\limits_{i=1}^{n}
(a_{i}-\bar{a})(b_{i}-\bar{b})}{[\sum\limits_{i=1}^{n}(a_{i}-\bar{a})^2\sum\limits_{i=1}^{n}(b_{i}-\bar{b})^2]^\frac{1}{2}}$
 \smallskip\\
\midrule
Description & The equation calculates Pearson correlation coefficients $\rho$ applied to two
datasets $A: \mathbb{R}^n$ and $B :\mathbb{R}^n$ both of size $n$.
 $\bar{a}$ and $\bar{b}$ are sample means (DD1) of $A$ and $B$, respectively.
              $\rho$ is the Pearson correlation coefficient between $A$ and $B$.
              Assumptions A1--A5 must hold when calculating this correlation.\\
  \bottomrule
\end{tabular}
\end{minipage}\\

\subsubsection{Coordinate Convention} \label{SecCoordConvention}

Section 4.a.iii of the SRS documents describes the coordinate system for the
fMRI images.  This information is necessary to make the requirements
unambiguous, since there are several choices for coordinate system for medical
images.  As an example, the SRS defines the Anatomical Coordinate System, which
describes the standard anatomical position of a human being using 3 orthogonal
planes: axial/transverse (plane parallel to the ground that separates the body
into head (superior) and tail (inferior) positions), coronal/frontal (plane
perpendicular to the ground that divides the body into front (anterior) and back
(posterior) positions), and sagittal/median (plane that divides the body into
right and left positions.  3dfim+ uses NIfTI data files
(\url{https://nifti.nimh.nih.gov/}) that store voxels from right to left to
create rows, rows from anterior to posterior to create slices and slices from
superior to inferior to create volumes.  This information was only partially
provided in the original documentation for 3dfim+.

\subsubsection{Rank Function} \label{SecRankFunction}

Calculating the Spearman and Quadrant correlation coefficients~\cite{Ward2000}
requires the use of the rank function.  The original documentation for 3dfim+
had incomplete documentation of the rank function.  The rank function is defined
in the SRS as a data definition.

The rank of data points is determined by sorting them in an ascending order and
assigning a value according to their position in the sorted list.  If ties
exist, the average of all of the tied positions is calculated as the rank.
Mathematically, the rank of
element $a$ in dataset $A$ is defined as follows:\\

\noindent $\mbox{rank}(a, A): \mathbb{R} \times \mathbb{R}^n \rightarrow \mathbb{R}$\newline
$\mbox{rank}(a, A) \equiv \mbox{avg}(\mbox{indexSet}(a, \mbox{sort}(A)))$\newline

\noindent $\mbox{indexSet}(a, B): \mathbb{R} \times \mathbb{R}^n \rightarrow \mbox{ set of }
\mathbb{N}$\newline
$\mbox{indexSet}(a, B) \equiv \{j: \mathbb{N} | j \in [1..|B|]
\wedge B_j = a : j \}$\newline

\noindent $\mbox{sort}(A): \mathbb{R}^n \rightarrow \mathbb{R}^n$\newline
$\mbox{sort}(A) \equiv B: \mathbb{R}^n, \mbox{ such that }$\newline
$\forall (a: \mathbb{R} | a \in A : \exists(b: \mathbb{R} | b \in B: b = a)
\wedge \mbox{count}(a, A) = \mbox{count}(b, B)) \wedge \forall (i: \mathbb{N} | i \in [1..|A|-1] : B_i \leq B_{i+1})$\newline

\noindent $\mbox{count}(a, A): \mathbb{R} \times \mathbb{R}^n \rightarrow \mathbb{N}$\newline
$\mbox{count}(a, A): + (x: \mathbb{N} | x \in A \wedge x = a : 1)$\newline

\noindent $\mbox{avg}(C): \mbox{ set of } \mathbb{N} \rightarrow \mathbb{R}$\newline
$\mbox{avg}(C) \equiv + (x: \mathbb{N} | x \in C : x) / |C|$\newline

The above equations use the Gries and Schneider
notation~\cite[p. 143]{GriesAndSchneider1993} for set building and evaluation of
an operator applied over a set of values.  Specifically, the expression
$(*x: X | R : P)$ means application of the operator $*$ to the values $P$ for
all $x$ of type $X$ for which range $R$ is true.  In the above equations, the
$*$ operators $\forall$, $\exists$ and $+$ are used.  Using this formal
notation, we can ensure that cases are not left out in the documentation.

\subsection{Test Cases} \label{SecTestCases}

To verify the implementation of 3dfim+, we developed test cases based on the
functional requirements documented in the SRS.  The results of the test cases
are used as evidence for goal GI (Figure~\ref{TopGoal}), which argues that the
implementation matches the SRS.  Since our case study is for SCS, verification
through testing is challenging.  The source of the challenge is that SCS differs
from most other software because the quantities of interest are continuous, as
opposed to discrete.  As shown for the calculation of the Pearson correlation
coefficient (Section~\ref{SecTheoryModel}), the inputs and outputs are
continuously valued real variables. Validating the requirements is difficult
because there are an infinite number of potential input values, many of which
cannot be represented as floating point numbers.  In general for SCS, the
correct value for the output variable is unknown.  That is, SCS problems
typically lack a test oracle~\cite{KellyEtAl2011}.  Fortunately for this MIA
example, 3dfim+, the correlation calculations are based on finite sets of real
numbers, so constructing a pseudo oracle using Matlab was relatively
straightforward.

We developed one Matlab test case per each functional requirement, to compare
their results with the results of 3dfim+. As an example, we had a test case to
check the correctness of the Pearson correlation coefficient, which is one of
the main functionalities of 3dfim+.  We used our Matlab pseudo oracle and AFNI
to visualize the results and obtain the indices of voxels.  Our input consisted
of 180 frames of 64$\times$64$\times$28 images.  In this test case, we found the
minimum and the maximum Pearson correlation coefficients and their locations.
For this test and others, we achieved the same results for both 3dfim+ and our
independently developed Matlab script.

However, the testing was not without its challenges.  Agreement between the
Matlab pseudo oracle and 3dfim+ took a considerable amount of time to achieve,
because the coordinate systems conventions for Matlab and AFNI are different.
Since this information was not documented in the original 3dfim+ manual, we were
unaware of this subtly.  The coordinate system description for 3dfim+ was added
to the SRS (as described in Section~\ref{SecCoordConvention}) after our
struggles with achieving test case agreement.

In the case of 3dfim+ a pseudo oracle was available.  For other MIA and SCS
software, other techniques may be needed for the verification that the
implementation matches the requirements~\cite{Smith2016}.  Where appropriate,
use can be made of the Method of Manufactured Solutions~\cite{Roache1998} and
metamorphic testing~\cite{KanewalaAndLundgren2016}.  For testing purposes, the
slower, but guaranteed correct, interval arithmetic~\cite{Hickey2001} can be
used to ensure that calculated answers lie within the guaranteed bounds.
Verification tests can also include plans for convergence studies.  The
discretization used in the numerical algorithm should be decreased (usually
halved) and the change in the solution assessed.  Although not used in the
current example, verification can also use non-testing techniques, such as code
walkthroughs, code inspections, and correctness proofs
etc.~\cite{GhezziEtAl2003, VanVliet2000}.

\subsection{Domain Expert Review}

An important piece of evidence for an assurance case is the domain expert
review.  Review of the SRS is important to reach a common understanding between
the software engineers and scientists.  As mentioned in the introduction
(Section~\ref{SecIntroduction}), building an assurance case facilitates bridging
the gap between software engineers and scientists.  The domain expert also
addresses the oracle problem outlined in the introduction.  Since the correct
answer is not known in general, an expert is needed to determine whether the
testing and other evidence is sufficient for building sufficient confidence in
the software.

Domain expert review appears in our assurance case as ``Domain experts/customers
approve the $<$quality$>$ of the documentation of the requirements.''  This
corresponds to GenericEvidence.2.3 in Figure~\ref{GenericEvidence}.  To ensure
our SRS is of high quality, a task-based inspection approach was
used~\cite{KellyAndShepard2000, Kelly2003}.  For the review process we assigned
a set of tasks asking questions about each section of the SRS.  We used Github
(\url{https://github.com/}) issue tracking for assigning the tasks and for
discussion.  Two sample review questions are reproduced below.

\begin{itemize}

\item[\textbf{Q7:}] Please let us know if all symbols in Theoretical Model T1
  (from Section~\ref{SecTheoryModel}) are defined. Is enough information
  provided that you could calculate the Pearson correlation coefficient if you
  are given datasets A and B.
\item[\textbf{Q10:}] Please let us know if Data Definition DD4 (Rank Function)
  (from Section~\ref{SecRankFunction}) is explained clearly or needs any
  additional information. Please let us know if the notation we are using for
  this function is clear and understandable.

\end {itemize}

A domain expert that completed the review for 3dfim+ has a degree in engineering
and over 10 years experience in medical imaging.  He therefore meets the
acceptance criteria given in E\_GenericEvidence.1 in Figure~\ref{TopGoal}.  The
reviewer went through all the assigned tasks and provided answers/suggestions.
For the most part, the SRS did not need to be modified as a result of the expert
review.  However, some of the symbols in the SRS, such as $\mathbb{N}$ for the
set of natural numbers, were clarified as a result of the discussion with the
expert reviewer.

The value of expert review is known in SCS.  For instance, the High Energy
Physics (HEP) software for the Compact Muon Solenoid (CMS) detector, which is
part of the Large Hadron Collider, has a GitHub pull request process that
involves automated testing, code quality checks and code review
(\url{http://cms-sw.github.io/PRWorkflow.html}.  This example is not unique in
SCS.  Applying such best practices contributes to building confidence.  However,
the reviews for the CMS software and for other SCS software are still often ad
hoc.  Questions like the following do not seem to be explicitly answered: What
are the reviewer's qualifications?  What software artifacts (code, documentation
and test cases) are they reviewing?  What issues/concerns are the reviewers
checking?  A review of the HEP community roadmap for future software and
computing research and development is completely silent on the idea of
formalizing the review process~\cite{StewartEtAl2017}.  Informal reviews are
certainly better than no reviews, but formal reviews have been demonstrated to
be more effective.  For instance, systematic code inspections of embedded
software have a defect removal effectiveness of 85\%~\cite{EbertAndJones2009}.  Assurance
cases answer the questions listed above; they make the review requirements
rigorous and defendable.  Moreover, building an assurance case, like the one for
3dfim+, shows explicitly that the code is not the only work product.  The
design, test plan, requirements etc., should also be reviewed.  As the domain
expert review highlights, the techniques for building confidence are already
employed in SCS; the shift to using assurance cases just means telling a more
complete and compelling story.

\section{Validation of Assurance Case Approach} \label{SecValidation}

The 3dfim+ case study provides evidence of the suitability of assurance cases
for SCS.  Although the original software was certainly built with care, problems
were still uncovered as a consequence of the systematic and rigorous process of
building a complete and defendable argument for correctness.  The need for
documentation, review and testing not only provides a means to improve the
software, it provides as a byproduct an explicit argument for correctness that
can be verified/replicated by third parties.

Before summarizing the assurance case improvements to 3dfim+, we should note
that we are not criticizing the original 3dfim+ software and its documentation.
The goal of the assurance case is to provide certifiable software, but the
original software did not have this goal.  It was written for researchers, not
for clinicians.  The users for 3dfim+ and readers of its documentation are
likely to be domain experts.  However, even for the existing audience for
3dfim+, there will likely be some novices.  The improvements noted below would
likely interest new users, since the new documentation is more complete and less
ambiguous than the original.  This benefit of improving software and its
documentation is also observed when retroactively writing an SRS for nuclear
safety analysis software~\cite{SmithAndKoothoor2016}.

Given the different audience that was envisioned, the original documentation
would not satisfy the GR goal (Figure~\ref{GRTop}) for high quality
requirements.  The existing documentation is not fully complete, unambiguous,
correct, consistent, verifiable, modifiable or traceable.  One of the main
ambiguities is through the absence of documentation on the coordinate system
(Section~\ref{SecCoordConvention}).  As mentioned previously, the absence of any
specification related to the coordinate system meant that comparing the 3dfim+
results to an independent calculation of the correlation was difficult.
Additional investigation was necessary to find the necessary details so that
both results were expressed in the same coordinate system.  Another ambiguity,
due to incomplete documentation, was for the definition of the rank function
(Section~\ref{SecRankFunction}).  In the original documentation the specific
definition of the rank function is not given.  This creates ambiguity when there
are ties in the data because there are multiple ways to deal with ties.  For
instance, all ties could be given the same rank, and then a gap could be left in
the ranking numbers.  Alternatively, ties could get the same rank, but no gap
could be included before listing the next ranking number.  Five different
ranking algorithms can be found on \url{https://en.wikipedia.org/wiki/Ranking}.
The one actually used by 3dfim+ gives the same ranking number to all ties, with
the rank being equal to the mean of what they would have under ordinal ranking.
This fact was not determined from the documentation, but by investigating the C
code implementation of 3dfim+.  The C code is the basis for the specification
given in Section~\ref{SecRankFunction}.

Although the software for 3dfim+ did not show any errors in its output, some
subtle concerns were raised by considering the assurance case GA for satisfying
the operational assumptions (Figure~\ref{AssumptionGoal}).  As shown in GA.1,
3dfim+ should not proceed if the input does not match the necessary assumptions.
However, the actual software does not check the input data.  GA.2 (``User is
aware of what inputs are valid'') is also not considered for 3dfim+.  The input
assumptions are not made explicit in the documentation and the user is not
warned that it is their responsibility to provide valid data.  As mentioned
previously, the user has the responsibility of determining whether the
statistical model used by 3dfim+ provides the right tool for them.  The need for
an explicit warning highlights a significant benefit of the assurance case
methodology - \emph{building an assurance case forces the developers to ask
  questions that they might not otherwise ask.}  We would likely not have
addressed ``how does a user know the inputs are valid?'' if the methodology had
not forced us to build an argument that the inputs satisfy the defined
operational assumptions (Figure~\ref{AssumptionGoal}).

Further evidence for the validity of applying assurance cases to SCS is the
success they have found for real time safety critical
systems~\cite{RinehartEtAl2015, Rushby2015, Wassyng2015}.  From the perspective
of assurance cases, SCS and real-time systems are not that different.  Both
domains require the qualities of correctness and reliability.  For qualities
where specific examples may differ in importance, such as the quality of
portability, the general approach of building an assurance case is the same, no
matter the quality.  The specific differences for different qualities will be in
terms of the arguments and the evidence, not in the overall approach.  In terms
of the relationship between developers, domain experts and regulators, the
motivating argument described at the beginning of this paper will usually apply
equally to SCS and real time software.  That is, in both cases, the regulators
will likely have less expert knowledge than the developers, which implies that
the case for building an argument for quality should be in the hands of the
developers.  If assurance cases are suitable for real-time systems, then they
are at least as suitable for SCS, since SCS is arguably simpler than real-time
software.  SCS does not generally have the same complexities in the external
environment, nor the same number of hazards, or concerns with emergent behaviour
from the interaction of multiple systems.

\section{Generalization of Approach and Future Work} \label{SecGeneralization}

Our example has focused on MIA, specifically 3dfim+, but generalizing the
assurance case approach to other SCS applications is straightforward.  Most of
the developed assurance case is not medical imaging specific, as illustrated by
the following points:

\begin{itemize}
\item The top level goal (Figure~\ref{TopGoal}) can be viewed as generic if it
  is parameterized by the software name.  That is, the name 3dfim+ could be
  replaced with any other SCS software application and the argument at this
  level would be unchanged.  This is not surprising as this top level itself was
  borrowed from an assurance case for medical devices~\cite{Wassyng2015}.
\item The context and assumptions for the top goal (Figure~\ref{TopAC}) would
  change for different SCS applications, but the type of questions to answer for
  the context would be similar, such as ``what is the intended functionality of
  the software?'', ``what is the intended software environment?'' etc.
\item The argument for the required qualities of the requirements is based on
  the IEEE standard, which allows decomposition of the quality concerns into
  arguments for correctness, completeness, consistency, unambiguity,
  verifiability, modifiability and traceability (Figure~\ref{GRTop}).  The IEEE
  standard is intended to apply to all software, not just medical imaging, or
  SCS.
\item The arguments for modifiability, generic evidence and operational
  assumptions (Figures~\ref{Modifiable},~\ref{GenericEvidence}
  and~\ref{AssumptionGoal}, respectively) could be used as a starting point for
  other examples.  The main difference will be in the required evidence.
\item The SRS template adopted for the requirements is not specific to MIA; it
  was developed for SCS in general and has been applied to cases such as thermal
  analysis of a nuclear fuel pin~\cite{SmithAndKoothoor2016}, mesh
  generation~\cite{SmithAndYu2009}, and
  others~\cite{SmithJegatheesanAndKelly2016}.
\end{itemize}

The need for building confidence in scientific software is not unique to medical
imaging analysis software, there are many cases where we need assurance, such as
for nuclear safety, computational medicine, climate modelling, etc.  The
verification of other scientific software, that provides the evidence at the
bottom of the assurance case, will likely vary from one problem to the next.
However, the tools and techniques for verification on scientific software
already exist; they do not need to be invented.  What is needed is a push to the
scientific software developers to use the existing techniques and document their
results so that others can build confidence in the software.  An expectation of
supplying an assurance case could provide this push.

Based on the current work and our review of past work on assurance cases, we
have identified a number of directions for the future development of assurance
cases, as follows:

\begin{itemize}
\item Additional Examples to Create a Template: As mentioned above, there are
  significant commonalities between SCS problems and predictable variabilities.
  Recording this information would make the creation of new assurance cases
  easier.  Since testing was relatively easy for 3dfim+, further exploration
  will be necessary for testing options, like the strategies mentioned in
  Section~\ref{SecTestCases}.
\item Build an Assurance Case for a Family of SCS Programs: Implicit in the
  previous discussion is that we have a single SCS program to build an assurance
  case, but in many situations, we are interested in a family of related SCS
  programs~\cite{SmithMcCutchanAndCao2007}, such as a family of linear solvers
  or ODE solvers.  In this case, we would investigate whether we should build an
  assurance case for the family, or a family of assurance cases?
\item Work on the Assurance Case from the Start: The current work produced the
  assurance case, SRS and test plan a posteriori.  As mentioned in
  Section~\ref{SecAssuranceCases}, assurance cases work best when they are used
  from the start of a project.  A particular benefit for research purposes will
  be a likely increase in the workload that can be assigned to domain reviewers,
  since they will likely have a greater vested interest in the success of the
  project than when it is a purely academic exercise.
\item Tool Support Improvement: Currently, there is no tool that provides an
  abstraction of goals and sub-goals to handle the complexity of the assurance
  case structure.  For instance, it would be nice to hide the details of a goal
  (or a context, justification, evidence or assumption) and only show the title.  This
  would improve readability.  Details could be revealed via clicking on the
  goals, or context etc.
\item Publishing Examples of Practical Assurance Cases: Currently, many existing
  assurance cases are not released due to proprietary rights. The more
  presentations on adoption of assurance cases and case studies, the better
  resources we have to learn about assurance cases.
\item Adding Formality to Assurance Cases: The means of expressing confidence in
  assurance cases and the top-level claims may benefit from further formality
  and rigour, as presented in~\cite{DiskinEtAl2018}. Adding formality could
  justify the completeness and consistency of claim decomposition and the
  credibility of the evidence.  A formal model will help, even if not all
  evidence will be mathematical.  The formal model will show the ideal situation
  and will clarify all of the requirements for a complete assurance case, even
  if some of the evidence itself has to be informal.
\end{itemize}

\section{Concluding Remarks} \label{SecConcludingRemarks}

This work has motivated assurance cases for SCS.  Assurance cases have already
been effectively used for safety cases for real time systems.  For SCS their
advantages include engaging domain experts, producing only necessary
documentation, and providing evidence that can potentially be
verified/replicated by a third party.  The engagement of the domain experts is
noteworthy because scientist end user developers have historically shown a
distrust of software engineering techniques and principles.  In particular, SCS
developers tend not to favour full documentation of requirements.  However,
their motivation should improve because an assurance case shows the necessity
and value of an SRS.  As more examples and tools become available, adoption of
assurance cases in SCS in general, and for the specific example of MIA, should
become more prevalent.  The FDA already strongly advises assurance cases for
newly developed infusion pumps~\cite{FDA2014}.

How to document an assurance case for SCS was illustrated via the MIA example of
the medical image analysis software, 3dfim+.  The 3dfim+ software analyzes
activity in the brain by computing the correlation between the measured and an
ideal brain signal.  This example was partly chosen because of recent concerns
about the validity of fMRI (Functional Magnetic Resonance Imaging) studies.  The
concerns centre around whether a parametric model is appropriate for fMRI data.
Although the software itself cannot determine whether it is the appropriate
model, the user should ask themselves this question.  The assurance case
highlighted how the user can be informed of the mathematical assumptions of the
model through the SRS, and the details of their responsibility through
in-program warnings.

The value of assurance cases for MIA was justified.  Although no errors were
found in the software outputs from 3dfim+, the exercise did highlight problems
with the original documentation and software.  The existing documentation was
shown to have ambiguities and omissions, such as an incompletely defined ranking
function and missing details on the coordinate system convention adopted.  In
addition, a potential concern for the software itself was identified.  As
mentioned above, running the software does not produce any warning about the
obligation of the user to provide data that matches the parametric statistical
model employed for the correlation calculations.  Further evidence for the
validity of applying assurance cases to SCS is the success they have found for
real time safety critical systems.  From the perspective of assurance cases, SCS
and real-time systems have much in common, with SCS generally having the
advantage of a operating within a simpler environment.

Our example has focused on MIA, but generalizing the assurance case approach to
other SCS applications is straightforward.  Each of the assurance case diagrams
shown had mostly generic content.  The main place where the examples become
specific are at the bottom of the argument, where the evidence is presented.
Although the evidence will be problem specific, the type of evidence needed,
like test reports, domain expert resumes, etc, will be similar between problems.

Although a concerted effort was made to make the assurance case for 3dfim+
convincing and complete, the specific argument for 3dfim+ is not the point of
this paper.  The important revelation about assurance cases is that they are for
communication, between experts, especially experts in different domains.  They
make what was previously implicit, explicit.  They force developers to ask
questions that they might not otherwise ask.  Is the evidence complete?  Do we
have an explicit argument that nothing important has been missed?  Much of the
needed evidence (test cases, expert reviews, etc) for an assurance case, with
the usual exception of requirements documentation, is already produced when
developing SCS.  The evidence is generally presented in an ad hoc way.  With
assurance cases, a third party does not have to use incomplete evidence to form
an opinion on the quality of a given work, they can use the full story, and
judge whether the story is convincing.  If a reviewer disagrees, they can point
to the portion of the argument that they feel is weak, and the developers will
have an opportunity to strengthen their argument.

The developed assurance case relies on the presence of requirements
documentation.  However, many in the SCS community believe that upfront
requirements are impossible, or at least infeasible~\cite{SmithEtAl2019}.  As a
consequence of this view, requirements are rarely explicitly recorded for SCS.
Can a convincing assurance case for correctness be produced without requirements
documentation as part of the evidence?  We do not believe so, since verification
exercises, like testing, require an explicit objective against which the results
can be judged.  However, we leave the challenge of producing a requirements
document free assurance case open to the SCS community.

\section{Acknowledgments} \label{Acknowledgments}

The assistance of Dr.\ Michael Noseworthy, from St.\ Joseph's hospital and
McMaster University, is gratefully acknowledged.  The 3dfim+ case study came
directly from his advice and fMRI demonstration.  Thanks also to Dr.\ Dean
Inglis for fulfilling the role of expert reviewer.


\begin{thebibliography}{52}


\ifx \showCODEN    \undefined \def \showCODEN     #1{\unskip}     \fi
\ifx \showDOI      \undefined \def \showDOI       #1{#1}\fi
\ifx \showISBNx    \undefined \def \showISBNx     #1{\unskip}     \fi
\ifx \showISBNxiii \undefined \def \showISBNxiii  #1{\unskip}     \fi
\ifx \showISSN     \undefined \def \showISSN      #1{\unskip}     \fi
\ifx \showLCCN     \undefined \def \showLCCN      #1{\unskip}     \fi
\ifx \shownote     \undefined \def \shownote      #1{#1}          \fi
\ifx \showarticletitle \undefined \def \showarticletitle #1{#1}   \fi
\ifx \showURL      \undefined \def \showURL       {\relax}        \fi
\providecommand\bibfield[2]{#2}
\providecommand\bibinfo[2]{#2}
\providecommand\natexlab[1]{#1}
\providecommand\showeprint[2][]{arXiv:#2}

\bibitem[\protect\citeauthoryear{Bishop and Bloofield}{Bishop and
  Bloofield}{1989}]%
        {Bishop98}
\bibfield{author}{\bibinfo{person}{Peter~G. Bishop} {and}
  \bibinfo{person}{Robin~E. Bloofield}.} \bibinfo{year}{1989}\natexlab{}.
\newblock \showarticletitle{A methodology for safety case development}. In
  \bibinfo{booktitle}{\emph{Industrial Perspectives of Safety-critical Systems:
  Proceedings of the Sixth Safety-critical Systems Symposium, Birmingham
  1998}}. \bibinfo{publisher}{Springer-Verlag}, \bibinfo{address}{London,
  {UK}}, \bibinfo{pages}{1--10}.
\newblock


\bibitem[\protect\citeauthoryear{Buckley, Davis, and Horch}{Buckley
  et~al\mbox{.}}{1993}]%
        {IEEESRS93}
\bibfield{author}{\bibinfo{person}{Fletcher~J.\ Buckley}, \bibinfo{person}{A.M.
  Davis}, {and} \bibinfo{person}{J.W. Horch}.} \bibinfo{year}{1993}\natexlab{}.
\newblock \bibinfo{booktitle}{\emph{{I}{E}{E}{E} Recommended Practice for
  Software Requirements Specifications}}.
\newblock \bibinfo{type}{{T}echnical {R}eport}. \bibinfo{institution}{The
  institute of Electrical and Electronics Engineers, Inc.},
  \bibinfo{address}{New York, {USA}}.
\newblock


\bibitem[\protect\citeauthoryear{Carver, Kendall, Squires, and Post}{Carver
  et~al\mbox{.}}{2007}]%
        {CarverEtAl2007}
\bibfield{author}{\bibinfo{person}{Jeffrey~C. Carver},
  \bibinfo{person}{Richard~P. Kendall}, \bibinfo{person}{Susan~E. Squires},
  {and} \bibinfo{person}{Douglass~E. Post}.} \bibinfo{year}{2007}\natexlab{}.
\newblock \showarticletitle{Software Development Environments for Scientific
  and Engineering Software: A Series of Case Studies}. In
  \bibinfo{booktitle}{\emph{ICSE '07: Proceedings of the 29th International
  Conference on Software Engineering}}. \bibinfo{publisher}{IEEE Computer
  Society}, \bibinfo{address}{Washington, DC, USA}, \bibinfo{pages}{550--559}.
\newblock
\showISBNx{0-7695-2828-7}
\urldef\tempurl%
\url{https://doi.org/10.1109/ICSE.2007.77}
\showDOI{\tempurl}


\bibitem[\protect\citeauthoryear{{CDRH}}{{CDRH}}{2002}]%
        {GPFDA}
\bibfield{author}{\bibinfo{person}{{CDRH}}.} \bibinfo{year}{2002}\natexlab{}.
\newblock \bibinfo{booktitle}{\emph{General Principles of Software Validation;
  Final Guidance for Industry and {FDA} Staff}}.
\newblock \bibinfo{type}{{T}echnical {R}eport}. \bibinfo{institution}{{US}
  Department Of Health and Human Services Food and Drug Administration {FDA},
  Center for Devices and Radiological {CDRH}, Health Center for Biologics
  Evaluation and Research}, \bibinfo{address}{Rockville, MD}.
\newblock


\bibitem[\protect\citeauthoryear{{Center for Devices and Radiological Health,
  CDRH}}{{Center for Devices and Radiological Health, CDRH}}{2002}]%
        {CDRH2002}
\bibfield{author}{\bibinfo{person}{{Center for Devices and Radiological Health,
  CDRH}}.} \bibinfo{year}{2002}\natexlab{}.
\newblock \bibinfo{booktitle}{\emph{General Principles of Software Validation;
  Final Guidance for Industry and {FDA} Staff}}.
\newblock \bibinfo{type}{{T}echnical {R}eport}. \bibinfo{institution}{{US}
  Department Of Health and Human Services Food and Drug Administration Center
  for Devices and Radiological Health Center for Biologics Evaluation and
  Research}, \bibinfo{address}{York, England}.
\newblock


\bibitem[\protect\citeauthoryear{Cleland, Sujan, Habli, and Medhurst}{Cleland
  et~al\mbox{.}}{2012}]%
        {HealthFoundation2012}
\bibfield{author}{\bibinfo{person}{G.M. Cleland}, \bibinfo{person}{M.A. Sujan},
  \bibinfo{person}{I. Habli}, {and} \bibinfo{person}{J. Medhurst}.}
  \bibinfo{year}{2012}\natexlab{}.
\newblock \bibinfo{booktitle}{\emph{Evidence: Using Safety Cases in Industry
  and Healthcare}}.
\newblock \bibinfo{publisher}{Health Foundation}, \bibinfo{address}{London}.
\newblock
\showISBNx{9781906461430}
\urldef\tempurl%
\url{https://books.google.ca/books?id=o8z-Ms9o3DMC}
\showURL{%
\tempurl}


\bibitem[\protect\citeauthoryear{Diskin, Maibaum, Wassyng, Wynn-Williams, and
  Lawford}{Diskin et~al\mbox{.}}{2018}]%
        {DiskinEtAl2018}
\bibfield{author}{\bibinfo{person}{Zinovy Diskin}, \bibinfo{person}{Tom
  Maibaum}, \bibinfo{person}{Alan Wassyng}, \bibinfo{person}{Stephen
  Wynn-Williams}, {and} \bibinfo{person}{Mark Lawford}.}
  \bibinfo{year}{2018}\natexlab{}.
\newblock \showarticletitle{Assurance via model transformations and their
  hierarchical refinement}. In \bibinfo{booktitle}{\emph{Proceedings of the
  21st International Conference on Models Driven Engineering Languages and
  Systems, MODELS 2018, Copenhagen, Denmark, October 14-19, 2018}}
  (2018-11-21). \bibinfo{publisher}{ACM}, \bibinfo{pages}{426 -- 436}.
\newblock
\urldef\tempurl%
\url{https://www.mcscert.ca/wp-content/uploads/2018/12/p426-diskin-1.pdf}
\showURL{%
\tempurl}


\bibitem[\protect\citeauthoryear{{Ebert} and {Jones}}{{Ebert} and
  {Jones}}{2009}]%
        {EbertAndJones2009}
\bibfield{author}{\bibinfo{person}{C. {Ebert}} {and} \bibinfo{person}{C.
  {Jones}}.} \bibinfo{year}{2009}\natexlab{}.
\newblock \showarticletitle{Embedded Software: Facts, Figures, and Future}.
\newblock \bibinfo{journal}{\emph{Computer}} \bibinfo{volume}{42},
  \bibinfo{number}{4} (\bibinfo{date}{April} \bibinfo{year}{2009}),
  \bibinfo{pages}{42--52}.
\newblock
\showISSN{0018-9162}
\urldef\tempurl%
\url{https://doi.org/10.1109/MC.2009.118}
\showDOI{\tempurl}


\bibitem[\protect\citeauthoryear{Eklunda, Nichols, and Knutssona}{Eklunda
  et~al\mbox{.}}{2016}]%
        {PNAS2016}
\bibfield{author}{\bibinfo{person}{Anders Eklunda}, \bibinfo{person}{Thomas
  Nichols}, {and} \bibinfo{person}{Hans Knutssona}.}
  \bibinfo{year}{2016}\natexlab{}.
\newblock \showarticletitle{A methodology for safety case development}.
\newblock \bibinfo{journal}{\emph{Proceedings of the National Academy of
  Sciences of the {U}nited {S}tates of America (PNAS)}} \bibinfo{volume}{113},
  \bibinfo{number}{28} (\bibinfo{year}{2016}), \bibinfo{pages}{7900--7905}.
\newblock


\bibitem[\protect\citeauthoryear{ESA}{ESA}{1991}]%
        {ESA1991}
\bibfield{author}{\bibinfo{person}{ESA}.} \bibinfo{year}{February
  1991}\natexlab{}.
\newblock \bibinfo{booktitle}{\emph{{ESA} Software Engineering Standards,
  {PSS-05-0} Issue 2}}.
\newblock \bibinfo{type}{{T}echnical {R}eport}. \bibinfo{institution}{European
  Space Agency}.
\newblock


\bibitem[\protect\citeauthoryear{Gade and Deshpande}{Gade and
  Deshpande}{2015}]%
        {GadeAndDeshpande2015}
\bibfield{author}{\bibinfo{person}{Dipak Gade} {and} \bibinfo{person}{Santosh
  Deshpande}.} \bibinfo{year}{2015}\natexlab{}.
\newblock \showarticletitle{Assurance Driven Software Design using Assurance
  Case Based Approach}.
\newblock \bibinfo{journal}{\emph{International Journal of Innovative Research
  in Computer and Communication Engineering}} \bibinfo{volume}{3},
  \bibinfo{number}{10} (\bibinfo{date}{October} \bibinfo{year}{2015}),
  \bibinfo{pages}{9121--9127}.
\newblock


\bibitem[\protect\citeauthoryear{Ghezzi, Jazayeri, and Mandrioli}{Ghezzi
  et~al\mbox{.}}{2003}]%
        {GhezziEtAl2003}
\bibfield{author}{\bibinfo{person}{Carlo Ghezzi}, \bibinfo{person}{Mehdi
  Jazayeri}, {and} \bibinfo{person}{Dino Mandrioli}.}
  \bibinfo{year}{2003}\natexlab{}.
\newblock \bibinfo{booktitle}{\emph{Fundamentals of Software Engineering}
  (\bibinfo{edition}{2nd} ed.)}.
\newblock \bibinfo{publisher}{Prentice Hall}, \bibinfo{address}{Upper Saddle
  River, NJ, USA}.
\newblock


\bibitem[\protect\citeauthoryear{Gries and Schneider}{Gries and
  Schneider}{1993}]%
        {GriesAndSchneider1993}
\bibfield{author}{\bibinfo{person}{David Gries} {and} \bibinfo{person}{Fred~B.
  Schneider}.} \bibinfo{year}{1993}\natexlab{}.
\newblock \bibinfo{booktitle}{\emph{A logical approach to discrete math}}.
\newblock \bibinfo{publisher}{Springer-Verlag Inc.}, \bibinfo{address}{New
  York}.
\newblock


\bibitem[\protect\citeauthoryear{Hatcliff, Heimdahl, Lawford, Maibaum, Wassyng,
  and Wurden}{Hatcliff et~al\mbox{.}}{2009}]%
        {HatcliffEtAl2009}
\bibfield{author}{\bibinfo{person}{John Hatcliff}, \bibinfo{person}{Mats
  Heimdahl}, \bibinfo{person}{Mark Lawford}, \bibinfo{person}{Tom Maibaum},
  \bibinfo{person}{Alan Wassyng}, {and} \bibinfo{person}{Fred Wurden}.}
  \bibinfo{year}{2009}\natexlab{}.
\newblock \showarticletitle{A Software Certification Consortium and its Top 9
  Hurdles}.
\newblock \bibinfo{journal}{\emph{Electronic Notes in Theoretical Computer
  Science}} \bibinfo{volume}{238}, \bibinfo{number}{4} (\bibinfo{year}{2009}),
  \bibinfo{pages}{11--17}.
\newblock
\showISSN{1571-0661}
\urldef\tempurl%
\url{https://doi.org/10.1016/j.entcs.2009.09.002}
\showDOI{\tempurl}


\bibitem[\protect\citeauthoryear{Hickey, Ju, and Van~Emden}{Hickey
  et~al\mbox{.}}{2001}]%
        {Hickey2001}
\bibfield{author}{\bibinfo{person}{Timothy Hickey}, \bibinfo{person}{Qun Ju},
  {and} \bibinfo{person}{Maarten~H. Van~Emden}.}
  \bibinfo{year}{2001}\natexlab{}.
\newblock \showarticletitle{Interval Arithmetic: From Principles to
  Implementation}.
\newblock \bibinfo{journal}{\emph{J. ACM}} \bibinfo{volume}{48},
  \bibinfo{number}{5} (\bibinfo{date}{Sept.} \bibinfo{year}{2001}),
  \bibinfo{pages}{1038--1068}.
\newblock
\showISSN{0004-5411}
\urldef\tempurl%
\url{https://doi.org/10.1145/502102.502106}
\showDOI{\tempurl}


\bibitem[\protect\citeauthoryear{IEEE}{IEEE}{1998}]%
        {IEEE1998}
\bibfield{author}{\bibinfo{person}{IEEE}.} \bibinfo{year}{1998}\natexlab{}.
\newblock \bibinfo{booktitle}{\emph{Recommended Practice for Software
  Requirements Specifications}}.
\newblock \bibinfo{type}{{T}echnical {R}eport} IEEE Std 830-1998.
  \bibinfo{institution}{The institute of Electrical and Electronics Engineers,
  Inc.} \bibinfo{pages}{1--40} pages.
\newblock
\urldef\tempurl%
\url{https://doi.org/10.1109/IEEESTD.1998.88286}
\showDOI{\tempurl}


\bibitem[\protect\citeauthoryear{Kanewala and Lundgren}{Kanewala and
  Lundgren}{2016}]%
        {KanewalaAndLundgren2016}
\bibfield{author}{\bibinfo{person}{Upulee Kanewala} {and}
  \bibinfo{person}{Anders Lundgren}.} \bibinfo{year}{2016}\natexlab{}.
\newblock \showarticletitle{Automated Metamorphic Testing of Scientific
  Software}.
\newblock In \bibinfo{booktitle}{\emph{Software Engineering for Science}},
  \bibfield{editor}{\bibinfo{person}{Jeffrey~C. Carver},
  \bibinfo{person}{Neil~Chue Hong}, {and} \bibinfo{person}{George
  Thiruvathukal}} (Eds.). \bibinfo{publisher}{Taylor \& Francis},
  \bibinfo{address}{Boca Raton, FL}, Chapter Examples of the Application of
  Traditional Software Engineering Practices to Science,
  \bibinfo{pages}{151--174}.
\newblock


\bibitem[\protect\citeauthoryear{Kelly and Shepard}{Kelly and Shepard}{2000}]%
        {KellyAndShepard2000}
\bibfield{author}{\bibinfo{person}{Diane Kelly} {and} \bibinfo{person}{Terry
  Shepard}.} \bibinfo{year}{2000}\natexlab{}.
\newblock \showarticletitle{Task-directed software inspection technique: an
  experiment and case study}. In \bibinfo{booktitle}{\emph{CASCON '00:
  Proceedings of the 2000 conference of the Centre for Advanced Studies on
  Collaborative research}}. \bibinfo{publisher}{IBM Press}, \bibinfo{pages}{6}.
\newblock
\urldef\tempurl%
\url{http://portal.acm.org/citation.cfm?id=782040#}
\showURL{%
\tempurl}


\bibitem[\protect\citeauthoryear{Kelly}{Kelly}{2007}]%
        {Kelly2007}
\bibfield{author}{\bibinfo{person}{Diane~F. Kelly}.}
  \bibinfo{year}{2007}\natexlab{}.
\newblock \showarticletitle{A Software Chasm: Software Engineering and
  Scientific Computing}.
\newblock \bibinfo{journal}{\emph{IEEE Software}} \bibinfo{volume}{24},
  \bibinfo{number}{6} (\bibinfo{year}{2007}), \bibinfo{pages}{120--119}.
\newblock
\showISSN{0740-7459}
\urldef\tempurl%
\url{https://doi.org/10.1109/MS.2007.155}
\showDOI{\tempurl}


\bibitem[\protect\citeauthoryear{Kelly, Smith, and Meng}{Kelly
  et~al\mbox{.}}{2011}]%
        {KellyEtAl2011}
\bibfield{author}{\bibinfo{person}{Diane~F. Kelly}, \bibinfo{person}{W.~Spencer
  Smith}, {and} \bibinfo{person}{Nicholas Meng}.}
  \bibinfo{year}{2011}\natexlab{}.
\newblock \showarticletitle{Software Engineering for Scientists}.
\newblock \bibinfo{journal}{\emph{Computing in Science \& Engineering}}
  \bibinfo{volume}{13}, \bibinfo{number}{5} (\bibinfo{date}{October}
  \bibinfo{year}{2011}), \bibinfo{pages}{7--11}.
\newblock


\bibitem[\protect\citeauthoryear{Kelly}{Kelly}{1999}]%
        {Kelly99}
\bibfield{author}{\bibinfo{person}{T.P. Kelly}.}
  \bibinfo{year}{1999}\natexlab{}.
\newblock \emph{\bibinfo{title}{Arguing Safety -- A Systematic Approach to
  Safety Case Management}}.
\newblock \bibinfo{thesistype}{Ph.D. Dissertation}. \bibinfo{school}{York
  University, Department of Computer Science Report {YCST}}.
\newblock


\bibitem[\protect\citeauthoryear{Kelly}{Kelly}{2003}]%
        {Kelly2003}
\bibfield{author}{\bibinfo{person}{Tim Kelly}.}
  \bibinfo{year}{2003}\natexlab{}.
\newblock \bibinfo{booktitle}{\emph{A Systematic Approach to Safety Case
  Management}}.
\newblock \bibinfo{type}{{T}echnical {R}eport} 04AE-149.
  \bibinfo{institution}{{SAE} International}.
\newblock


\bibitem[\protect\citeauthoryear{NASA}{NASA}{1989}]%
        {NASA1989}
\bibfield{author}{\bibinfo{person}{NASA}.} \bibinfo{year}{1989}\natexlab{}.
\newblock \bibinfo{booktitle}{\emph{Software requirements {DID},
  {SMAP-DID-P200-SW}, Release 4.3}}.
\newblock \bibinfo{type}{{T}echnical {R}eport}. \bibinfo{institution}{National
  Aeronautics and Space Agency}.
\newblock


\bibitem[\protect\citeauthoryear{Nejad}{Nejad}{2017}]%
        {Nejad2017}
\bibfield{author}{\bibinfo{person}{Mojdeh~Sayari Nejad}.}
  \bibinfo{year}{2017}\natexlab{}.
\newblock \emph{\bibinfo{title}{A Case Study in Assurance Case Development for
  Scientific Software}}.
\newblock \bibinfo{thesistype}{Master's\ thesis}. \bibinfo{school}{McMaster
  University}, \bibinfo{address}{Hamilton, ON, Canada}.
\newblock
\newblock
\shownote{\url{http://hdl.handle.net/11375/23075}.}


\bibitem[\protect\citeauthoryear{Northrop}{Northrop}{2004}]%
        {Northrop2004}
\bibfield{author}{\bibinfo{person}{L. Northrop}.}
  \bibinfo{year}{2004}\natexlab{}.
\newblock \bibinfo{title}{Achieving Product Qualities Through Software
  Architecture Practices}.
\newblock
\newblock
\urldef\tempurl%
\url{http://www.sei.cmu.edu/architecture/cseet04.pdf}
\showURL{%
\tempurl}


\bibitem[\protect\citeauthoryear{{Office for Nuclear Regulation}}{{Office for
  Nuclear Regulation}}{2016}]%
        {ONR2016}
\bibfield{author}{\bibinfo{person}{{Office for Nuclear Regulation}}.}
  \bibinfo{year}{2016}\natexlab{}.
\newblock \bibinfo{title}{A guide to Nuclear Regulation in the {UK}}.
\newblock
\newblock
\urldef\tempurl%
\url{http://www.onr.org.uk/documents/a-guide-to-nuclear-regulation-in-the-uk.pdf}
\showURL{%
\tempurl}


\bibitem[\protect\citeauthoryear{Parnas and Clements}{Parnas and
  Clements}{1986}]%
        {ParnasAndClements1986}
\bibfield{author}{\bibinfo{person}{David~L. Parnas} {and} \bibinfo{person}{P.C.
  Clements}.} \bibinfo{year}{1986}\natexlab{}.
\newblock \showarticletitle{A Rational Design Process: How and Why to Fake it}.
\newblock \bibinfo{journal}{\emph{IEEE Transactions on Software Engineering}}
  \bibinfo{volume}{12}, \bibinfo{number}{2} (\bibinfo{date}{February}
  \bibinfo{year}{1986}), \bibinfo{pages}{251--257}.
\newblock


\bibitem[\protect\citeauthoryear{Pernet and Nichols}{Pernet and
  Nichols}{2016}]%
        {Guardian2016}
\bibfield{author}{\bibinfo{person}{Cyril Pernet} {and} \bibinfo{person}{Tom
  Nichols}.} \bibinfo{year}{2016}\natexlab{}.
\newblock \bibinfo{title}{Has a software bug really called decades of brain
  imaging research into question?}
\newblock
\newblock
\urldef\tempurl%
\url{https://www.theguardian.com/science/head-quarters/2016/sep/30/}
\showURL{%
\tempurl}


\bibitem[\protect\citeauthoryear{Rinehart, Knight, and Rowanhill}{Rinehart
  et~al\mbox{.}}{2015}]%
        {RinehartEtAl2015}
\bibfield{author}{\bibinfo{person}{David~J. Rinehart}, \bibinfo{person}{John~C.
  Knight}, {and} \bibinfo{person}{Jonathan Rowanhill}.}
  \bibinfo{year}{2015}\natexlab{}.
\newblock \bibinfo{booktitle}{\emph{Current Practices in Constructing and
  Evaluating Assurance Cases with Applications to Aviation}}.
\newblock \bibinfo{type}{{T}echnical {R}eport} CR-2014-218678.
  \bibinfo{institution}{National Aeronautics and Space Administration (NASA)},
  \bibinfo{address}{Langley Research Centre, Hampton, Virginia}.
\newblock


\bibitem[\protect\citeauthoryear{Roache}{Roache}{1998}]%
        {Roache1998}
\bibfield{author}{\bibinfo{person}{Patrick~J. Roache}.}
  \bibinfo{year}{1998}\natexlab{}.
\newblock \bibinfo{booktitle}{\emph{Verification and Validation in
  Computational Science and Engineering}}.
\newblock \bibinfo{publisher}{Hermosa Publishers},
  \bibinfo{address}{Albuquerque, New Mexico}.
\newblock


\bibitem[\protect\citeauthoryear{Rushby}{Rushby}{2015}]%
        {Rushby2015}
\bibfield{author}{\bibinfo{person}{John Rushby}.}
  \bibinfo{year}{2015}\natexlab{}.
\newblock \bibinfo{booktitle}{\emph{The Interpretation and Evaluation of
  Assurance Cases}}.
\newblock \bibinfo{type}{{T}echnical {R}eport} SRI-CSL-15-01.
  \bibinfo{institution}{Computer Science Laboratory, SRI International},
  \bibinfo{address}{Menlo Park, CA}.
\newblock
\newblock
\shownote{Available at
  \url{http://www.csl.sri.com/users/rushby/papers/sri-csl-15-1-assurance-cases.pdf}.}


\bibitem[\protect\citeauthoryear{Segal}{Segal}{2005}]%
        {Segal2005}
\bibfield{author}{\bibinfo{person}{Judith Segal}.}
  \bibinfo{year}{2005}\natexlab{}.
\newblock \showarticletitle{When Software Engineers Met Research Scientists: A
  Case Study}.
\newblock \bibinfo{journal}{\emph{Empirical Software Engineering}}
  \bibinfo{volume}{10}, \bibinfo{number}{4} (\bibinfo{date}{October}
  \bibinfo{year}{2005}), \bibinfo{pages}{517--536}.
\newblock
\showISSN{1382-3256}
\urldef\tempurl%
\url{https://doi.org/10.1007/s10664-005-3865-y}
\showDOI{\tempurl}


\bibitem[\protect\citeauthoryear{Segal}{Segal}{2008}]%
        {Segal2008}
\bibfield{author}{\bibinfo{person}{Judith Segal}.}
  \bibinfo{year}{2008}\natexlab{}.
\newblock \showarticletitle{Models of Scientific Software Development}. In
  \bibinfo{booktitle}{\emph{Proceedings of the First International Workshop on
  Software Engineering for Computational Science and Engineering (SECSE
  2008)}}. In conjunction with the 30th International Conference on Software
  Engineering (ICSE), \bibinfo{publisher}{ACM}, \bibinfo{address}{Leipzig,
  Germany}, \bibinfo{pages}{1--6}.
\newblock
\urldef\tempurl%
\url{http://www.cse.msstate.edu/~SECSE08/schedule.htm}
\showURL{%
\tempurl}


\bibitem[\protect\citeauthoryear{Segal and Morris}{Segal and Morris}{2008}]%
        {SegalAndMorris2008}
\bibfield{author}{\bibinfo{person}{Judith Segal} {and} \bibinfo{person}{Chris
  Morris}.} \bibinfo{year}{2008}\natexlab{}.
\newblock \showarticletitle{Developing Scientific Software}.
\newblock \bibinfo{journal}{\emph{IEEE Software}} \bibinfo{volume}{25},
  \bibinfo{number}{4} (\bibinfo{date}{July/August} \bibinfo{year}{2008}),
  \bibinfo{pages}{18--20}.
\newblock


\bibitem[\protect\citeauthoryear{Smith, Srinivasan, and Shankar}{Smith
  et~al\mbox{.}}{2019}]%
        {SmithEtAl2019}
\bibfield{author}{\bibinfo{person}{Spencer Smith}, \bibinfo{person}{Malavika
  Srinivasan}, {and} \bibinfo{person}{Sumanth Shankar}.}
  \bibinfo{year}{2019}\natexlab{}.
\newblock \showarticletitle{Debunking the Myth that Upfront Requirements are
  Infeasible for Scientific Computing Software}. In
  \bibinfo{booktitle}{\emph{2019 International Workshop on Software Engineering
  for Science (held in conjunction with ICSE'19)}}. \bibinfo{pages}{1--8}.
\newblock


\bibitem[\protect\citeauthoryear{Smith}{Smith}{2016}]%
        {Smith2016}
\bibfield{author}{\bibinfo{person}{W.~Spencer Smith}.}
  \bibinfo{year}{2016}\natexlab{}.
\newblock \showarticletitle{A Rational Document Driven Design Process for
  Scientific Computing Software}.
\newblock In \bibinfo{booktitle}{\emph{Software Engineering for Science}},
  \bibfield{editor}{\bibinfo{person}{Jeffrey~C. Carver},
  \bibinfo{person}{Neil~Chue Hong}, {and} \bibinfo{person}{George
  Thiruvathukal}} (Eds.). \bibinfo{publisher}{Taylor \& Francis},
  \bibinfo{address}{Boca Raton, FL}, Chapter Examples of the Application of
  Traditional Software Engineering Practices to Science,
  \bibinfo{pages}{33--63}.
\newblock


\bibitem[\protect\citeauthoryear{Smith, Jegatheesan, and Kelly}{Smith
  et~al\mbox{.}}{2016}]%
        {SmithJegatheesanAndKelly2016}
\bibfield{author}{\bibinfo{person}{W.~Spencer Smith}, \bibinfo{person}{Thulasi
  Jegatheesan}, {and} \bibinfo{person}{Diane~F. Kelly}.}
  \bibinfo{year}{2016}\natexlab{}.
\newblock \showarticletitle{Advantages, Disadvantages and Misunderstandings
  About Document Driven Design for Scientific Software}. In
  \bibinfo{booktitle}{\emph{Proceedings of the Fourth International Workshop on
  Software Engineering for High Performance Computing in Computational Science
  and Engineering (SE-HPCCE)}}.
\newblock
\newblock
\shownote{8 pp.}


\bibitem[\protect\citeauthoryear{Smith and Koothoor}{Smith and
  Koothoor}{2016}]%
        {SmithAndKoothoor2016}
\bibfield{author}{\bibinfo{person}{W.~Spencer Smith} {and}
  \bibinfo{person}{Nirmitha Koothoor}.} \bibinfo{year}{2016}\natexlab{}.
\newblock \showarticletitle{A Document-Driven Method for Certifying Scientific
  Computing Software for Use in Nuclear Safety Analysis}.
\newblock \bibinfo{journal}{\emph{Nuclear Engineering and Technology}}
  \bibinfo{volume}{48}, \bibinfo{number}{2} (\bibinfo{date}{April}
  \bibinfo{year}{2016}), \bibinfo{pages}{404--418}.
\newblock
\showISSN{1738-5733}
\urldef\tempurl%
\url{https://doi.org/10.1016/j.net.2015.11.008}
\showDOI{\tempurl}


\bibitem[\protect\citeauthoryear{Smith and Lai}{Smith and Lai}{2005}]%
        {SmithAndLai2005}
\bibfield{author}{\bibinfo{person}{W.~Spencer Smith} {and} \bibinfo{person}{Lei
  Lai}.} \bibinfo{year}{2005}\natexlab{}.
\newblock \showarticletitle{A New Requirements Template for Scientific
  Computing}. In \bibinfo{booktitle}{\emph{Proceedings of the First
  International Workshop on Situational Requirements Engineering Processes --
  Methods, Techniques and Tools to Support Situation-Specific Requirements
  Engineering Processes, SREP'05}},
  \bibfield{editor}{\bibinfo{person}{J.~Ralyt\'{e}},
  \bibinfo{person}{P.~\.{A}gerfalk}, {and} \bibinfo{person}{N.~Kraiem}} (Eds.).
  In conjunction with 13th IEEE International Requirements Engineering
  Conference, \bibinfo{publisher}{IEEE}, \bibinfo{address}{Paris, France},
  \bibinfo{pages}{107--121}.
\newblock


\bibitem[\protect\citeauthoryear{Smith, Lai, and Khedri}{Smith
  et~al\mbox{.}}{2007a}]%
        {SmithEtAl2007}
\bibfield{author}{\bibinfo{person}{W.~Spencer Smith}, \bibinfo{person}{Lei
  Lai}, {and} \bibinfo{person}{Ridha Khedri}.}
  \bibinfo{year}{2007}\natexlab{a}.
\newblock \showarticletitle{Requirements Analysis for Engineering Computation:
  A Systematic Approach for Improving Software Reliability}.
\newblock \bibinfo{journal}{\emph{Reliable Computing, Special Issue on Reliable
  Engineering Computation}} \bibinfo{volume}{13}, \bibinfo{number}{1}
  (\bibinfo{date}{February} \bibinfo{year}{2007}), \bibinfo{pages}{83--107}.
\newblock


\bibitem[\protect\citeauthoryear{Smith, McCutchan, and Cao}{Smith
  et~al\mbox{.}}{2007b}]%
        {SmithMcCutchanAndCao2007}
\bibfield{author}{\bibinfo{person}{W.~Spencer Smith}, \bibinfo{person}{John
  McCutchan}, {and} \bibinfo{person}{Fang Cao}.}
  \bibinfo{year}{2007}\natexlab{b}.
\newblock \showarticletitle{Program Families in Scientific Computing}. In
  \bibinfo{booktitle}{\emph{7$^{th}$ OOPSLA Workshop on Domain Specific
  Modelling ({DSM}'07)}}, \bibfield{editor}{\bibinfo{person}{Jonathan
  Sprinkle}, \bibinfo{person}{Jeff Gray}, \bibinfo{person}{Matti Rossi}, {and}
  \bibinfo{person}{Juha-Pekka Tolvanen}} (Eds.).
  \bibinfo{address}{Montr\'{e}al, Qu\'{e}bec}, \bibinfo{pages}{39--47}.
\newblock


\bibitem[\protect\citeauthoryear{Smith, Nejad, and Wassyng}{Smith
  et~al\mbox{.}}{2018}]%
        {SmithEtAl2018_ICSEPoster}
\bibfield{author}{\bibinfo{person}{W.~Spencer Smith},
  \bibinfo{person}{Mojdeh~Sayari Nejad}, {and} \bibinfo{person}{Alan Wassyng}.}
  \bibinfo{year}{2018}\natexlab{}.
\newblock \showarticletitle{Assurance Cases for Scientific Computing Software
  (poster)}. In \bibinfo{booktitle}{\emph{ICSE 2018 Proceedings of the 40th
  International Conference on Software Engineering}}.
  \bibinfo{publisher}{Association for Computing Machinery {(ACM)}},
  \bibinfo{address}{Gothenburg, Sweden}, \bibinfo{pages}{1--2}.
\newblock
\newblock
\shownote{2 pp.}


\bibitem[\protect\citeauthoryear{Smith and Yu}{Smith and Yu}{2009}]%
        {SmithAndYu2009}
\bibfield{author}{\bibinfo{person}{W.~Spencer Smith} {and} \bibinfo{person}{Wen
  Yu}.} \bibinfo{year}{2009}\natexlab{}.
\newblock \showarticletitle{A Document Driven Methodology for Improving the
  Quality of a Parallel Mesh Generation Toolbox}.
\newblock \bibinfo{journal}{\emph{Advances in Engineering Software}}
  \bibinfo{volume}{40}, \bibinfo{number}{11} (\bibinfo{date}{Nov.}
  \bibinfo{year}{2009}), \bibinfo{pages}{1155--1167}.
\newblock
\urldef\tempurl%
\url{https://doi.org/10.1016/j.advengsoft.2009.05.003}
\showDOI{\tempurl}


\bibitem[\protect\citeauthoryear{Spriggs}{Spriggs}{2012}]%
        {Spriggs2012}
\bibfield{author}{\bibinfo{person}{John Spriggs}.}
  \bibinfo{year}{2012}\natexlab{}.
\newblock \bibinfo{booktitle}{\emph{{GSN} - The Goal Structuring Notation, A
  Structured Approach to Presenting Arguments}}.
\newblock \bibinfo{publisher}{Springer}, \bibinfo{address}{Hayling Island,
  {UK}}.
\newblock
\showISBNx{978-1-4471-2311-8}
\urldef\tempurl%
\url{https://doi.org/10.1007/978-1-4471-2312-5}
\showDOI{\tempurl}


\bibitem[\protect\citeauthoryear{Stewart et~al\mbox{.}}{Stewart
  et~al\mbox{.}}{2017}]%
        {StewartEtAl2017}
\bibfield{author}{\bibinfo{person}{Graeme Stewart} {et~al\mbox{.}}}
  \bibinfo{year}{2017}\natexlab{}.
\newblock \showarticletitle{{A Roadmap for HEP Software and Computing R\&D for
  the 2020s}}.
\newblock \bibinfo{journal}{\emph{arXiv}} (\bibinfo{year}{2017}).
\newblock
\showeprint[arxiv]{physics.comp-ph/1712.06982}


\bibitem[\protect\citeauthoryear{Storer}{Storer}{2017}]%
        {Storer2017}
\bibfield{author}{\bibinfo{person}{Tim Storer}.}
  \bibinfo{year}{2017}\natexlab{}.
\newblock \showarticletitle{Bridging the Chasm: A Survey of Software
  Engineering Practice in Scientific Programming}.
\newblock \bibinfo{journal}{\emph{ACM Comput. Surv.}} \bibinfo{volume}{50},
  \bibinfo{number}{4}, Article \bibinfo{articleno}{47} (\bibinfo{date}{Aug.}
  \bibinfo{year}{2017}), \bibinfo{numpages}{32}~pages.
\newblock
\showISSN{0360-0300}
\urldef\tempurl%
\url{https://doi.org/10.1145/3084225}
\showDOI{\tempurl}


\bibitem[\protect\citeauthoryear{{U.S. Food and Drug Administration}}{{U.S.
  Food and Drug Administration}}{2014}]%
        {FDA2014}
\bibfield{author}{\bibinfo{person}{{U.S. Food and Drug Administration}}.}
  \bibinfo{year}{2014}\natexlab{}.
\newblock \bibinfo{title}{Infusion Pumps Total Product Life Cycle: Guidance for
  Industry and FDA Staff}.
\newblock \bibinfo{howpublished}{on-line}.
\newblock


\bibitem[\protect\citeauthoryear{van Vliet}{van Vliet}{2000}]%
        {VanVliet2000}
\bibfield{author}{\bibinfo{person}{Hans van Vliet}.}
  \bibinfo{year}{2000}\natexlab{}.
\newblock \bibinfo{booktitle}{\emph{Software Engineering (2nd ed.): Principles
  and Practice}}.
\newblock \bibinfo{publisher}{John Wiley \& Sons, Inc.}, \bibinfo{address}{New
  York, NY, USA}.
\newblock
\showISBNx{0-471-97508-7}


\bibitem[\protect\citeauthoryear{Ward}{Ward}{2000}]%
        {Ward2000}
\bibfield{author}{\bibinfo{person}{B.~Douglas Ward}.}
  \bibinfo{year}{2000}\natexlab{}.
\newblock \bibinfo{booktitle}{\emph{Program 3dfim+}}.
\newblock Biophysics Research Institute, Medical College of Wisconsin.
\newblock


\bibitem[\protect\citeauthoryear{Wassyng, Singh, Geven, Proscia, Wang, Lawford,
  and Maibaum}{Wassyng et~al\mbox{.}}{2015}]%
        {Wassyng2015}
\bibfield{author}{\bibinfo{person}{Alan Wassyng}, \bibinfo{person}{Neeraj~Kumar
  Singh}, \bibinfo{person}{Mischa Geven}, \bibinfo{person}{Nicholas Proscia},
  \bibinfo{person}{Hao Wang}, \bibinfo{person}{Mark Lawford}, {and}
  \bibinfo{person}{Tom Maibaum}.} \bibinfo{year}{2015}\natexlab{}.
\newblock \showarticletitle{Can Product-Specific Assurance Case Templates Be
  Used as Medical Device Standards?}
\newblock \bibinfo{journal}{\emph{{IEEE} Design {\&} Test}}
  \bibinfo{volume}{32}, \bibinfo{number}{5} (\bibinfo{year}{2015}),
  \bibinfo{pages}{45--55}.
\newblock
\urldef\tempurl%
\url{https://doi.org/10.1109/MDAT.2015.2462720}
\showDOI{\tempurl}


\bibitem[\protect\citeauthoryear{Weinstock, Lipson, and Goodenough}{Weinstock
  et~al\mbox{.}}{2007}]%
        {Charles2007}
\bibfield{author}{\bibinfo{person}{Charles~B. Weinstock},
  \bibinfo{person}{Howard~F. Lipson}, {and} \bibinfo{person}{John Goodenough}.}
  \bibinfo{year}{2007}\natexlab{}.
\newblock \bibinfo{booktitle}{\emph{Arguing Security -- Creating Security
  Assurance Cases}}.
\newblock \bibinfo{type}{{T}echnical {R}eport}. \bibinfo{institution}{Software
  Engineering Institute Carnegie Mellon University}, \bibinfo{address}{4500
  Fifth Avenue, Pittsburgh, {PA}}.
\newblock
\urldef\tempurl%
\url{http://resources.sei.cmu.edu/asset_files/WhitePaper/2013_019_001_293637.pdf}
\showURL{%
\tempurl}


\bibitem[\protect\citeauthoryear{Zowghi and Gervasi}{Zowghi and
  Gervasi}{2013}]%
        {ThreeCs}
\bibfield{author}{\bibinfo{person}{Didar Zowghi} {and}
  \bibinfo{person}{Vincenzo Gervasi}.} \bibinfo{year}{2013}\natexlab{}.
\newblock \bibinfo{booktitle}{\emph{The Three {Cs} of Requirements:
  Consistency, Completeness, and Correctness}}.
\newblock \bibinfo{type}{{T}echnical {R}eport}. \bibinfo{institution}{Faculty
  of Information Technology University of Technology, Sydney , Australia}.
\newblock


\end{thebibliography}


\end{document}